\documentclass[aps,prb,final,twocolumn,showpacs,amsmath,amssymb,superscriptaddress]{revtex4-1}
\pdfoutput=1
\usepackage{hyperref}
\usepackage{graphicx}
\usepackage{aurical}
\usepackage[T1]{fontenc}
\usepackage{color}

\DeclareGraphicsExtensions{.pdf}

\newcommand{\br}{\mathbf{r}}
\newcommand{\barbr}{\bar {\mathbf r}}
\newcommand{\barbarbr}{\bar{\bar {\mathbf r}}}
\newcommand{\dif}{\mathrm{d}}
\newcommand{\SC}{\mathcal{S}}

\newcommand{\VT}{V_{\rm T}}
\newcommand{\EF}{\varepsilon_{\rm F}}
\newcommand{\kF}{k_{\rm F}}
\newcommand{\comment}[1]{}

\newcommand{\ipcms}{Institut de Physique et de Chimie des Mat{\'e}riaux de 
Strasbourg, Universit\'e de Strasbourg, CNRS UMR 7504,\\
23 rue du Loess, BP 43, F-67034 Strasbourg Cedex 2, France}
\newcommand{\augsburg}{Universit\"at Augsburg, Institut f\"ur Physik, Theoretische Physik II,
D-86135 Augsburg, Germany}
\newcommand{\pullman}{Department of Physics and Astronomy, PO Box 642814, 
Washington State University, Pullman, WA 99164-2814, USA}

\def\bepsilon{\bar{\varepsilon}}
\def\bbepsilon{\bar {\bar{\varepsilon}}}
\def\bk{\bar{k}}
\def\bbk{\bar {\bar{k}}}
\def\ba{\bar{a}}
\def\bba{\bar {\bar{a}}}
\def\bl{\bar{l}}
\def\bbl{\bar {\bar{l}}}
\def\tepsilon{\varepsilon^{(\mathrm{t})}}
\def\Me{M_{\rm e}}
\def\coola{\mbox{\Fontauri A}} % the fancy font for the little a matrix elements
\def\r]{\right]}
\def\l[{\left[}
\def\hbj{\hat{\mbox{\bf \j}}}

\begin{document}

\title{Theory of scanning gate microscopy}

\author{Cosimo Gorini}
\affiliation{\ipcms}
\affiliation{\augsburg}
\author{Rodolfo A.\ Jalabert}
\affiliation{\ipcms}
\author{Wojciech Szewc}
\affiliation{\ipcms}
\author{Steven Tomsovic}
\affiliation{\pullman}
\author{Dietmar Weinmann}
\affiliation{\ipcms}

\date{\today}

%\pacs{85.35.Ds, %Quantum interference devices
%     07.79.-v, %Scanning probe microscopes and components
%     73.23.-b, %Electronic transport in mesoscopic systems
%     72.10.-d  %Theory of electronic transport; scattering mechanisms
%}

\begin{abstract}
A systematic theory of the conductance measurements of non-invasive (weak probe)
scanning gate microscopy is presented that provides an interpretation of
what precisely is being measured. A scattering approach is used to
derive explicit expressions for the first and second order conductance
changes due to the perturbation by the tip potential in terms of the
scattering states of the unperturbed structure. In the case of a
quantum point contact, the first order correction dominates at the
conductance steps and vanishes on the plateaus where the second order
term dominates.  Both corrections are non-local for a generic structure.
 Only in special cases, such as that of a centrally symmetric quantum
point contact in the conductance quantization regime,  can the second
order correction be unambiguously related with the local current
density. In the case of an abrupt quantum point contact we are able to
obtain analytic expressions for the scattering eigenfunctions and thus
evaluate the resulting conductance corrections.
\end{abstract}

\maketitle

\section{Introduction}
\label{intro}

Understanding the underlying mechanisms of the electronic transport properties of nanostructures 
in the quantum coherent regime is of great fundamental interest, and also of foremost 
importance for possible applications in nanoelectronics devices and quantum computing.
The scanning gate microscopy (SGM) technique, born more than a decade ago,\cite{eriksson96} 
provides additional information about coherent transport beyond 
that obtained in traditional transport experiments, and has therefore attracted considerable interest. 

In SGM, the transport through a nanostructure is measured while the charged tip of an atomic 
force microscope (AFM) is scanned over the sample. The tip-induced potential scatters 
electrons and thereby influences the sample's transport properties. 
One then studies how transport coefficients, for example the linear 
conductance through the nanostructure, depend on the position of the tip over the sample. 
The technique (for a brief review see Ref.~[\onlinecite{sellier11}]) has been applied to 
quantum point contacts (QPCs) defined in two-dimensional electron gases (2DEGs),  
\cite{topinka00a,topinka01a,aidala07,jura07,jura09a,jura10,schnez11b,aoki12,kozikov12} 
and to a variety of other systems, including carbon nanotubes,\cite{bachtold00} 
quantum dots fabricated in various systems (semiconductor 2DEGs,\cite{pioda04,fallahi05} 
carbon nanotubes,\cite{woodside02} semiconductor nanowires,\cite{bleszynski07} and 
graphene \cite{schnez11a}), small Aharonov-Bohm rings,\cite{hackens06,martins07a,pala08a} 
Hall bars,\cite{baumgartner07} edge channels,\cite{paradiso10,paradiso11,paradiso12}
quantum billiards,\cite{crook03} and bilayer graphene.\cite{connolly12}

Recently, SGM has been extended to so-called scanning probe microscopy \cite{Boyd_West}, 
where the scanning of a charged tip together with the measurements
of the resulting Coulomb blockade peaks energy-shifts of a
one-dimensional quantum dot enables the extraction of the electronic
wave-functions density profile.

Numerical calculations of the conductance changes as a function of the tip position in 
the non-perturbative case yield conductance maps closely related to the local current flow.\cite{aidala07,leroy05}
Already in the first SGM measurements on QPCs\cite{topinka00a,topinka01a}
the idea was advanced that the conductance maps image the current flow.
Tip-induced features, such as interference fringes, were observed for QPC-tip distances 
below the phase coherence length, but larger than a thermal length. 
Those patterns were attributed to the 
interference between electron paths scattered back to the QPC by the tip, 
and by an impurity within a thermal length of the tip.\cite{heller05} 
Such interference patterns disappear for very clean
samples\cite{jura07} except at very low temperatures,\cite{jura09a} where the 
thermal length exceeds the QPC-tip distance.  Interestingly, the pattern for a QPC
tuned to the first conductance plateau has been observed to be rather different from 
the one obtained when the QPC is tuned to the first step.\cite{jura09a}  The spacial
periodicity of the interferences when the tip moves away from the QPC
was expected to be half a Fermi wave-length, assuming it would be due to
the interference between the electron path scattered back by the tip
with another one that does not visit the tip. Though this is consistent
with most experimental observations, recent experiments show different
periodicity in different regions of the same sample\cite{kozikov12} 
and challenge this simple picture.

In the presence of strong AFM probes,
the interpretation of SGM data in terms of electron flow appears consistent with numerical 
studies of the local current density and the SGM response in QPCs \cite{metalidis05,cresti06,leroy05,kramer08}.
On the other hand, SGM experiments on small ring structures combined with numerical simulations \cite{martins07a,pala08a} 
have concluded that in this situation the conductance change is related to the local density of states 
rather than to the local current. 
Though a generally applicable theory leading to an unambiguous and quantitative interpretation 
of the SGM data is still lacking, experiments continue to propose novel and interesting uses of 
the SGM technique. For example, while theoretical modeling suggested looking for interaction 
effects in the regime where the tip is very close to the QPC,\cite{freyn08a} experimental evidence 
indicates that SGM in the nonlinear transport voltage regime gives access to information about
electron-electron scattering inside the sample.\cite{jura10}

In spite of the numerous measurements and theoretical investigations 
to be found in the literature, it is not always obvious how 
the now considerable quantity of data is to be interpreted. 
Further theoretical studies of coherent transport in the presence of a local scatterer 
are needed to fill in the gaps of current understanding.

In a previous paper,\cite{JSTW_2010} we initiated a systematic approach to the theory of the 
conductance change induced by a weak local perturbation. The resulting general expressions allow 
one to calculate the correction to the conductance to first and second order in the 
perturbation caused by the tip, starting from the scattering properties and wave-functions 
of the unperturbed nanostructure. 
It was found that the first order correction is suppressed when the QPC is tuned to a conductance plateau. 
This explains a change in the SGM response depending on the tuning of the QPC, 
and is consistent with a thermal enhancement of the interference patterns proposed for a 
QPC tuned close to the edge of a plateau \cite{abbout11}.
Most of the experiments performed up to date use a very strong tip-induced potential 
to obtain significant contrast.  Such a regime most likely requires
higher order corrections to be considered or a non-perturbative approach.
However, within the goal of investigating the electronic properties of the unperturbed sample, 
a perturbative approach appears as the first step towards the understanding of what is measured by non-invasive SGM,
and experiments relying on weaker probes would also be desirable -- as required, for example,
for the recently proposed scanning probe microscopy setup\cite{Boyd_West}.

This paper provides a detailed derivation of the expressions presented in
Ref.~[\onlinecite{JSTW_2010}], and addresses the question of the relationship between the SGM signal and 
local quantities, such as the current density, by applying the general formulas to the case of the SGM 
response of a QPC with abrupt openings. This model is capable of reproducing
quite accurately and in a controlled way the features related to the QPC conductance quantization,
and, crucially, allows one to obtain expressions for the scattering wave-functions, 
thus making it possible to compare the SGM response to the local current density. These 
two quantities are directly related only if: (i) the QPC is tuned to a conductance plateau,
and (ii) the symmetry of the structure is such that the wave leaving the QPC can be written 
as the product of a radial and an angular part, as it is for an outgoing wave from a point-like 
source in a clean medium. 
However, when the symmetry of the structure is broken 
by imperfections of the sample or by disorder in the neighboring 2DEG, 
even on a plateau the SGM response is no longer given by the local current density.   

The paper is organized as follows. Section~\ref{formalism} presents the scattering 
formalism used to evaluate the SGM response. The derivation of the first and second order conductance 
corrections follows in Secs.~\ref{sec_1st} and \ref{sec_2nd}, respectively. The calculation of the 
scattering wave-functions and the scattering eigenmodes for an abrupt QPC is presented in 
Secs.~\ref{sec_QPC} and \ref{sec_TEQPC}. The symmetry of the unperturbed structure is shown 
in Sec.~\ref{sec_discussion} to play a key role for the determination of the conductance corrections, 
and the example of an abrupt QPC is used for the discussion of the connection between conductance 
corrections and current densities. A summary is presented in the concluding Sec.\ \ref{sec_conclusions}. 
Details of the energy integrations used in Secs.~\ref{sec_1st} and \ref{sec_2nd}, alternative routes to reach the main results, 
the comparison of our expressions with exact results available for one-dimensional models, and that with 
the formalism proposed in Ref.~[\onlinecite{gasparian96}] all appear in the appendices.

\section{Model description and scattering formalism}
\label{formalism}

The conduction electrons in a typical SGM setup are described by a Hamiltonian
\begin{equation}
\label{eq:h0plusvT}
H=H_0+\VT \ ,
\end{equation}
where $H_0$ represents the unperturbed structure to be characterized and $\VT$ the 
electrostatic potential generated by the perturbing tip. While most of the analysis is developed 
for an arbitrary $H_0$, the case of a quantum point contact is treated in detail. Similarly, 
general results for a non-specified $\VT$ are established, and then the case of local spatial perturbations is the focus. 

The scattering theory of quantum conductance~\cite{FisherLee,buttiker88,jalabert00,mello04,LB} 
assumes that the unperturbed 
structure, considered as a non-interacting scatterer, is connected to electron reservoirs through 
disorder-free leads of finite cross section and semi-infinite in the longitudinal direction. Even 
though the scattering theory is applicable to an arbitrary number of leads and spatial dimensions, 
we will restrict ourselves to the case of interest of a two-lead setup within a two-dimensional 
space spanned by vectors $\br=(x,y)$.

The electrons in the reservoirs are assumed to be free, and therefore their dispersion relation is
$\varepsilon=\hbar^2 k^{2}/2\Me$, with $k$ the magnitude of the two-dimensional wave-vector and 
$\Me$ the effective electron mass. Taking the $x$-direction as the longitudinal one, the incoming $(-)$ 
and outgoing $(+)$ {\it modes} in lead 1 (left) and 2 (right) with energy $\varepsilon$ can be taken, 
respectively, as
\begin{subequations}
\label{allleadst} 
\begin{equation}
\varphi_{1,\varepsilon,a}^{(\mp)}(\br)  =  
\frac{c}{\sqrt{k_{a}}} \ \exp{[\pm i k_{a}^\mp x]}
\ \phi_{a}(y) \ , \quad x <  \ 0 \ ,
\end{equation}
\begin{equation}
\varphi_{2,\varepsilon,a}^{(\mp)}(\br)  =  
\frac{c}{\sqrt{k_{a}}} \ \exp{[\mp i k_{a}^\mp x]} 
\ \phi_{a}(y) \ , \quad x >  \ 0 \ .
\end{equation}
\end{subequations}
With $\phi_{a}(y)$ we denote the wave-function of the $a$\textsuperscript{th} transverse {\it channel} 
(with quantized energy $\varepsilon^{(\mathrm{t})}_a$), whereas $k_{a}$ is the longitudinal wave-vector, 
$v_{a}=\hbar k_{a}/\Me$ the longitudinal velocity, and 
$\varepsilon^{(\mathrm{l})}_a = \hbar^2 k^{2}_{a}/2\Me$ the longitudinal energy. 
The relationship $\varepsilon=\varepsilon^{(\mathrm{t})}_a + \varepsilon^{(\mathrm{l})}_a$ determines, 
for each lead, the $N$ propagating modes which satisfy $k_{a}^2 > 0$ at energy $\varepsilon$. 
We always take $k_{a} > 0$. The notation $k_{a}^\mp$ stands for an infinitesimal negative (positive) 
imaginary part given to $k_a$ for incoming (outgoing) modes. Choosing the constant $c$ equal to $1$ 
amounts to the so-called unit-flux normalization for the lead modes. We adopt however a slightly 
different convention and set
\begin{equation}
\label{eq:normal}
c=\sqrt{\frac{\Me}{2\pi \hbar^2}} \ .
\end{equation}
This is notationally simpler and leads to a current 
density in the $x$-direction and a current (per spin and unit energy) associated with the 
right-(left-)moving mode $1(2),\varepsilon,a$ given by $(\pm e/h) |\phi_{a}(y)|^2$ and 
$\pm e/h$, respectively. The overall signs result from our convention of taking as positive the current of 
positive charges moving from left to right.

Although any separable potential can be treated if $k_a$ is taken as $x$-dependent,
we consider a confining potential that is $x$-independent in the asymptotic regions. 
For simplicity 
we choose a hard wall confinement in the $y$ direction by taking leads of width $2W$. Then
\begin{equation}\label{eq:transverse_energy_lead}
\varepsilon^{(\mathrm{t})}_a = 
\frac{\hbar^2q_a^2 }{2\Me}\
\end{equation}
and
\begin{equation}
\label{eq:phi_transv}
\phi_{a}(y) = \frac{(-1)^p}{\sqrt{W}} 
\sin{\left[q_a(y-W)\right]} \, ,
\end{equation}
where $q_a=\pi a/2W$ is the transverse wave-vector satisfying $k_a=\sqrt{k^2-q_a^2}$, and 
$p={\rm Int}\{a/2\}$. The introduction of the phase $(-1)^p$ is a matter of convenience for writing 
the scattering wave-functions, as done in Sec.~\ref{sec_2nd}. It merely gives a sign alternation within each 
of the families of even and odd (in $y$) modes corresponding, respectively, to odd and even $a$. 

Once a quantum coherent scatterer (of linear extension $L$ in the $x$ direction) is placed at the 
coordinate origin, the incoming modes 
$\varphi_{1(2),\varepsilon,a}^{(-)}$ give rise to {\it outgoing scattering states} (defined for all $x$) that in the asymptotic regions are, respectively,
\begin{widetext}
\begin{subequations}\label{allscats}
 % ref the whole group by \eqref{allscats}; ref single eq. by e.g. \eqref{scatst1}
\begin{align}
\Psi_{1,\varepsilon,a}^{(0)}(\br) 
&= 
\left\lbrace 
\begin{array}{ll}
\varphi_{1,\varepsilon,a}^{(-)}(\br) + \sum_{b=1}^{N} r_{ba} \, \varphi_{1,\varepsilon,b}^{(+)}(\br),
 & x \ll -L/2 \\
\sum_{b=1}^{N} t_{ba} \, \varphi_{2,\varepsilon,b}^{(+)}(\br), & x \gg L/2  
\end{array} 
\right. \label{scatst1} 
\\
\Psi_{2,\varepsilon,a}^{(0)}(\br) 
&= 
\left\lbrace 
\begin{array}{ll}
\varphi_{2,\varepsilon,a}^{(-)}(\br) + \sum_{b=1}^{N} r^{\prime}_{ba} \, 
\varphi_{2,\varepsilon,b}^{(+)}(\br), & x \gg L/2 \\
\sum_{b=1}^{N} t^{\prime}_{ba} \, \varphi_{1,\varepsilon,b}^{(+)}(\br), & x \ll -L/2 
\end{array} 
\right. \label{scatst2} 
\end{align}
\end{subequations}
\end{widetext}
In order to simplify the notation we do not write the standard $+$ label corresponding to outgoing 
scattering states. (The incoming scattering states generated from the outgoing modes $\varphi_{1(2),\varepsilon,a}^{(+)}$ 
will not be used in this work.) The index $(0)$ is used for labeling unperturbed quantities depending on $H_0$ only. 
The $N \times N$ matrices $r$ ($r'$) and $t$ ($t'$) characterize, respectively, the reflection and 
transmission matrices from lead $1\ (2)$. The $2N \! \times \! 2N$ scattering matrix $S$, relating 
incoming and outgoing modes, is given by
\begin{equation}
\label{eq:scatt_mat}
S = \left( \begin{array}{cc}
r & t' \\
t & r'
\end{array} \right) \, .
\end{equation}
Current conservation dictates the unitarity of the scattering matrix ($S S^{\dagger} = I$). 
In the absence of magnetic fields, which is the situation considered in this work, 
time-reversal symmetry implies that $S$ is a symmetric matrix ($S^{\mathrm{T}}=S$).
For simplicity we do not explicitly write the energy dependence of the various components of the 
scattering matrix, nor do we include the index $(0)$ that should in principle be assigned for 
consistency of the notation. 

The normalization (\ref{eq:normal}) ensures that the scattering states constitute an 
orthonormal basis verifying
\begin{equation}
\label{state_norm}
\int \dif {\br} \ \Psi_{l,\varepsilon,a}^{(0)*}(\br) \
\Psi_{\bar{l},\bar{\varepsilon},\bar{a}}^{(0)}(\br) = 
\delta_{l \bar{l}} \ \delta(\varepsilon-\bar{\varepsilon}) \ 
\delta_{a \bar{a}} \ ,
\end{equation}
and have the same flux normalization as the lead states.

The transmission and reflection amplitudes between modes $a$ and $b$ can be obtained, 
respectively, from the retarded Green function $\mathcal{G}^{(0)}(\br,\bar{\br},\varepsilon)$
associated to the unperturbed Hamiltonian $H_0$ [\onlinecite{FisherLee}]
\begin{widetext}
\begin{subequations}
\label{allTRAMs}
\begin{eqnarray}
\hspace{1.0cm}
t_{ba} & = & i\hbar(v_{a}v_{b})^{1/2} \ 
\exp{\left[-i(k_b^{+} x - k_a^{+} {\bar x})\right]}
\int_{\SC_{x}} \dif y \int_{\SC_{\bar x}} \dif {\bar y} \ \phi_{b}^{*}(y) \ 
\mathcal{G}^{(0)}(\br,\bar{\br},\varepsilon)
\ \phi_{a}({\bar y}) \ ,
\label{eq:TRAM0} \\
\hspace{1.0cm} r_{ba} & = & -\delta_{ab} \
\exp{\left[i(k_b^{+} x + k_a^{+} {\bar x})\right]} \exp{\left[ik_b^{+}|x-{\bar x}|\right]}
\nonumber \\
\displaystyle
& + &  i \hbar(v_{a}v_{b})^{1/2}
\ \exp{\left[-i(k_b^{+} x + k_a^{+} {\bar x})\right]} 
\int_{\SC_{x}} \dif y \int_{\SC_{\bar x}} 
\dif {\bar y} \ \phi_{b}^{*}(y) \ 
\mathcal{G}^{(0)}(\br,\bar{\br},\varepsilon)
\ \phi_{a}({\bar y})
\ .
\label{eq:TRAM1}
\end{eqnarray}
\end{subequations}
\end{widetext}
The integrations take place at the transverse cross sections ${\cal S}_{\bar x}$ on the left 
lead and ${\cal S}_x$ on the right (left) lead for the transmission (reflection) amplitudes.
Since only retarded Green functions are used, the standard $(+)$ label for $\mathcal{G}^{(0)}$ is 
dropped. 

The Landauer-B\"uttiker approach to quantum transport derives the conductance of a coherent 
scatterer taking as building blocks the current carried by the scattering states,\cite{LB} 
in accordance with the fact the in the traditional conductance measurement setups only the 
total integrated currents are relevant.
However, the SGM technique alters the current density by introducing additional carrier backscattering, thus providing a spatial resolution that yields information about the scatterer beyond that of 
the traditional setup. In order to understand the outcome of SGM measurements, and the perturbative
approach that is developed, it is useful to reformulate the scattering formalism starting from
the current operator, defined as 
\begin{equation}
\hbj(\br)=\frac{e}{2\Me}\left[\hat{\mathbf{p}}\,
\delta(\hat{\br}-\br)+\delta(\hat{\br}-\br)\hat{\mathbf{p}}\right],
\end{equation}
where $\hat{\br}$ and $\hat{\mathbf{p}}$ denote the position and momentum operators, respectively.
The matrix elements of the $x$-component of the current density in the scattering states basis read 
\begin{widetext}
\begin{equation}\label{j0matrix}
\l[j^{x}(\br)\r]_{\ba a}^{\bl l}(\bar{\varepsilon},\varepsilon)=
\frac{e\hbar}{2i\Me}
\l[
\Psi_{\bar{l},\bar{\varepsilon},\bar{a}}^{(0)*}(\br) 
\ \frac{\partial}{\partial x}\Psi_{l,\varepsilon,a}^{(0)}(\br) -
\Psi_{l,\varepsilon,a}^{(0)}(\br) \ \frac{\partial}{\partial x}\Psi_{\bar{l},
\bar{\varepsilon},\bar{a}}^{(0)*}(\br) 
\r].
\end{equation}
\end{widetext}
The diagonal matrix element, that we write $j^{x(0)}_{l,\varepsilon, a}(\br)$, represents the 
current density per spin and unit energy associated with the state $\Psi_{l,\varepsilon,a}^{(0)}$. 
For a given incoming lead $l$ and energy $\varepsilon$ it is useful to define 
an $N \times N$ current matrix ${\cal I}^{(0)}_{l,\varepsilon}$ whose elements 
in the scattering states subspace $l,\varepsilon$ are given by 
\begin{equation}
\l[{\cal I}^{(0)}_{l,\varepsilon}\r]_{\ba a} = \int_{{\cal S}_x}\,\dif\,y\,
\l[j^{x(0)}(\br)\r]_{\ba a}^{l l}(\varepsilon,\varepsilon) \ .
\end{equation}
The independence of ${\cal I}_{1,\varepsilon}^{(0)}$ from the cross section ${\cal S}_x$ 
chosen for the integration is a consequence of current conservation. 
Given the one-to-one correspondence between incoming modes and outgoing scattering states, 
and the asymptotic form \eqref{allscats} of the latter, 
the current matrix elements (involving scattering states) can be identified 
with those of $t^\dagger t$ (involving lead modes), that is, 
\begin{equation}
\label{tt_vs_I}
\left[{\cal I}_{1,\varepsilon}^{(0)}\right]_{\ba a}=\frac{e}{h}\left[t^{\dagger}t\right]_{\ba a}.
\end{equation}
The diagonal matrix element $\l[{\cal I}^{(0)}_{1,\varepsilon}\r]_{a a}$
is the current (per spin and unit energy) associated with the scattering state $1,\varepsilon,a$
\begin{equation}
\label{eq:I0state}
I^{(0)}_{1,\varepsilon,a} = 
\frac{e}{h} \sum_{b=1}^{N} |t_{ba}|^2
= \frac{e}{h} 
\left(1-\sum_{b=1}^{N} |r_{ba}|^2\right) \ .
\end{equation}
The total current from left to right can be written as
\begin{equation}
\label{eq:I0}
I^{(0)}_1 = \int_{\mu_2}^{\mu_1} \dif \varepsilon \sum_{a=1}^{N} 2\pi\hbar v_{a} \rho_{a}(\varepsilon) \ 
I^{(0)}_{1,\varepsilon,a} \, ,
\end{equation}
with $\rho_{a}(\varepsilon)=(\pi \hbar v_a)^{-1}$ the 
one-dimensional density of lead modes (including the spin degeneracy factor) and $\mu_{1(2)}$ the 
electrochemical potential in the left (right) reservoir. 

Working in linear response to the applied bias $eV=\mu_1-\mu_2$ leads to the two-probe 
Landauer-B\"uttiker formula for the dimensionless conductance of the scatterer \cite{LB} 
\begin{equation}
\label{eq:g0}
g^{(0)}= \frac{1}{2 e^2/h} \frac{I^{(0)}_1}{V} = \mathrm{Tr}[t^\dagger t],
\end{equation} 
where all quantities are calculated at the Fermi energy $\varepsilon_{\rm F}$ of the reservoirs, 
and the trace is taken over the incoming, right-moving modes. 

Within the approach to be developed, it is convenient to write the symmetric scattering 
matrix $S$ in the polar decomposition \cite{jalabert00,mello04}
\begin{equation}
S = \left( \begin{array}{cc}
u_{1}^\mathrm{T}		& \ 0	\\
0		& \ u_{2}^\mathrm{T}
\end{array} \right)
\left( \begin{array}{cc}
-{\cal R}	& \hspace{0.5cm} {\cal T}	\\
{\cal T}	& \hspace{0.5cm} {\cal R}
\end{array} \right)
\left( \begin{array}{cc}
u_{1}	& \ 0	\\
0	& \ u_{2}
\end{array} \right) \ ,
\label{eq:Spol}
\end{equation}
where $u_{1(2)}$ are $N \times N$ unitary matrices, while ${\cal R}$ and ${\cal T}$ are diagonal 
matrices defined by the {\it reflection} and {\it transmission eigenvalues}, respectively given by
\begin{subequations}
\label{allrtcals}
\begin{align}
{\cal R}_m &= \left(\frac{\lambda_m}{1+\lambda_m}\right)^{1/2} \ ,
\label{rtcala}
\\
{\cal T}_m &= \left(\frac{1}{1+\lambda_m}\right)^{1/2} \ ,
\label{rtcalb}
\end{align}
\end{subequations}
with $\lambda_m$ real positive. 

Since $t^\dagger t = u_{1}^{\dagger}{\cal T}^2u_{1}$ and 
$t^{\prime \dagger} t^{\prime} = u_{2}^{\dagger}{\cal T}^2u_{2}$, the matrices $u_{1(2)}$ diagonalize, 
respectively, $t^\dagger t$ and $t^{\prime \dagger} t^{\prime}$. The transmission eigenvectors, 
or {\it transmission eigenmodes}, are of the form
\begin{subequations}
\label{allleadsteigen} 
\begin{align}
\varrho_{1,\varepsilon,m}^{(-)}(\br) &=
\sum_{a=1}^{N} \left[u_{1}\right]_{m a}^{*} \
\varphi_{1,\varepsilon,a}^{(-)}(\br)
 \ , \quad x <  \ 0 \ ,
\\
\varrho_{2,\varepsilon,m}^{(-)}(\br) &=
\sum_{a=1}^{N} \left[u_{2}\right]_{m a}^{*} \
\varphi_{2,\varepsilon,a}^{(-)}(\br)
\ , \quad x >  \ 0 \ .
\end{align}
\end{subequations}
The dimensionless conductance only depends of the transmission eigenvalues ${\cal T}_m$ as
\begin{equation}
g^{(0)} = \sum_{m=1}^{N} \ {\cal T}_m^2 \, .
\label{eq:geigen}
\end{equation} 
Our goal is to determine the change of the conductance from $g^{(0)}$ to $g$ when the full 
Hamiltonian $H$ of Eq.~(\ref{eq:h0plusvT}) is considered. In the perturbative approach
in powers of $\VT$ developed in Secs.~\ref{sec_1st} and \ref{sec_2nd}, 
the properties of the unperturbed system are important, particularly the eigenvalues and 
eigenfunctions of the current matrix ${\cal I}_{1,\varepsilon}^{(0)}$.
In Secs.~\ref{sec_QPC} and \ref{sec_TEQPC} we will focus on QPCs, building such eigenfunctions and 
then studying in Sec.~\ref{sec_discussion} the relation between the unperturbed current densities 
and the tip-induced conductance changes.

\section{First-order perturbation in the tip potential}
\label{sec_1st}

In order to analyze the effect of the tip voltage in its least invasive
form we will treat the potential $\VT$ of Eq.~(\ref{eq:h0plusvT}) as
a perturbation to $H_0$. 
Two substantially equivalent approaches are possible.
One can compute the corrections to the retarded Green function
of the unperturbed structure $\mathcal{G}^{(0)}$ via the Dyson equation,
thus obtaining the modified transmission and reflection amplitudes through Eqs.~\eqref{allTRAMs}.
Alternatively, the corrections to the unperturbed scattering states $\Psi_{l,\varepsilon,a}^{(0)}(\br)$
can be computed by means of the Lippmann-Schwinger equation,
and the modified states can then be employed to obtain the corresponding current density and full current.
We mentioned that the computation of the local current densities is fundamental for our purposes,
as these need be compared to the SGM-induced conductance corrections in order to establish
the relation -- if any -- between the two quantities. 
We will therefore follow the second route, pursuing further the reasoning of the previous Section. 
A discussion of the Green function approach is given however in Appendix~\ref{app_green}.
The Lippmann-Schwinger equation relates the unperturbed scattering state 
$\Psi_{l,\varepsilon,a}^{(0)}(\br)$ to the corresponding perturbed state 
$\Psi_{l,\varepsilon,a}(\br)$ according to
\begin{equation}\label{eq:Lippmann-Schwinger}
\Psi_{l,\varepsilon,a}(\br)=\Psi_{l,\varepsilon,a}^{(0)}(\br)+
\int \dif \bar{\br} \ \mathcal{G}^{(0)}(\br,\bar{\br},\varepsilon)  
\VT(\bar{\br})  \Psi_{l,\varepsilon,a}(\bar{\br}) \ .
\end{equation}
The Green function $\mathcal{G}^{(0)}$, introduced in Eq.~(\ref{allTRAMs}),
has the following spectral decomposition in the scattering states basis
\begin{equation}
\label{eq:G0}
\mathcal{G}^{(0)}(\br,\bar{\br},\varepsilon) = \sum_{\bar{l}=1}^{2} 
\int_{\varepsilon^{(\mathrm{t})}_1}^{\infty} 
\frac{\dif\bar{\varepsilon}}{\varepsilon^{+}-\bar{\varepsilon}}
\sum_{\bar{a}=1}^{\bar{N}} \ \Psi_{\bar{l},\bar{\varepsilon},\bar{a}}^{(0)*}(\bar{\br}) 
\Psi_{\bar{l},\bar{\varepsilon},\bar{a}}^{(0)}(\br) \ ,
\end{equation} 
having defined $\varepsilon^{\pm}=\varepsilon \pm i \eta$, with $\eta$ positive infinitesimal.

In Eq.~(\ref{eq:Lippmann-Schwinger}) the first order correction in $\VT$ (Born approximation) to 
the scattering state is obtained as
\begin{equation}\label{eq:Born_approx}
\Psi_{l,\varepsilon,a}^{(1)}(\br)=
\int \dif \bar{\br} \ \mathcal{G}^{(0)}(\br,\bar{\br},\varepsilon)  
\VT(\bar{\br})  \Psi_{l,\varepsilon,a}^{(0)}(\bar{\br}) \ .
\end{equation}
The corresponding change of the current density associated with the state $l,\varepsilon,a$ is given by
\begin{widetext}
\begin{equation}
\label{eq:j1}
j_{l,\varepsilon,a}^{x(1)}(\br) = 2\sum_{\bar{l}=1}^{2} \mathrm{Re}\left\lbrace 
\int_{\varepsilon^{(\mathrm{t})}_1}^{\infty} 
\frac{\dif\bar{\varepsilon}}{\varepsilon^{+}-\bar{\varepsilon}} \
\sum_{\bar{a}=1}^{\bar{N}} 
\left[j^{x}(\br)\right]_{a \ba}^{l \bl}(\varepsilon,\bar{\varepsilon}) \
\left[\VT\right]^{\bl l}_{\ba a}(\bar{\varepsilon},\varepsilon) 
\right\rbrace \, ,
\end{equation}
where the matrix elements of the perturbing potential in the scattering states basis are 
\begin{equation}
\label{eq:meVT}
\left[\VT\right]^{\bl l}_{\ba a}(\bar{\varepsilon},\varepsilon) =
\int \dif \br \
\Psi_{\bar{l},\bar{\varepsilon},\bar{a}}^{(0)*}(\br) \ \VT(\br) \ 
\Psi_{l,\varepsilon,a}^{(0)}(\br)\, . 
\end{equation}
The change of the current associated with the scattering state $1,\varepsilon,a$, obtained by 
integrating Eq.\ (\ref{eq:j1}) over a cross section $\SC_x$ in the right lead, is
\begin{equation}
I_{1,\varepsilon,a}^{(1)}(x) = \frac{e\hbar}{M} \ \sum_{\bar{l}=1}^{2}
\ \mathrm{Re}\left\lbrace\int_{\varepsilon^{(\mathrm{t})}_1}^{\infty} 
\frac{\dif\bar{\varepsilon}}{\varepsilon^{+}-\bar{\varepsilon}}
\sum_{\bar{a}=1}^{\bar{N}} 
Z^{1 \bl}_{a \ba}(\varepsilon,\bar{\varepsilon}) \ 
\left[\VT\right]_{\ba a}^{\bl 1}
(\bar{\varepsilon},\varepsilon)
\right\rbrace  \, .
\label{eq:deltaI}
\end{equation}
The final result is obviously independent of $x$ and the chosen lead (by current conservation).
In Eq.\ (\ref{eq:deltaI}) we have introduced a shorthand notation for quantities involving 
the unperturbed states
\begin{subequations}
\label{eq:allZ_def}
\begin{eqnarray}
\label{eq:Z_defa}
Z^{11}_{a \ba}(\varepsilon,\bar{\varepsilon}) &=& c^2 \sum_{b=1}^{\hat{N}}
\left(\sqrt{\frac{\bar{k}_b}{k_b}}+
\sqrt{\frac{k_b}{\bar{k}_b}}\right) 
t^{*}_{ba}{\bar t}^{\phantom{*}}_{b\bar{a}}
\exp{\left[i(\bar{k}^{+}_b-k^-_b)x\right] } \, ,  \\
Z^{12}_{a \ba}(\varepsilon,\bar{\varepsilon}) &=& 
c^2 \left\lbrace
\left(\sqrt{\frac{k_{\bar{a}}}{\bar{k}_{\bar{a}}}}-
\sqrt{\frac{\bar{k}_{\bar{a}}}{k_{\bar{a}}}}\right)
t^{*}_{\bar{a}a}\exp{\left[-i(\bar{k}^{-}_{\bar{a}}+k^-_{\bar{a}})x\right ] }
+ \sum_{b=1}^{\hat{N}}
\left(\sqrt{\frac{\bar{k}_b}{k_b}}+
\sqrt{\frac{k_b}{\bar{k}_b}}\right)t^{*}_{ba}
{\bar r}'_{b\bar{a}}
\exp{\left[i(\bar{k}^{+}_b-k^-_b)x\right]}
\right\rbrace , \nonumber
\\
\label{eq:Z_defb}
\end{eqnarray}
\end{subequations}
\end{widetext}
where the barred longitudinal wave-vectors and scattering amplitudes are taken at the total energy 
$\bar{\varepsilon}$, and $\hat{N}= \mathrm{min}\lbrace N,\bar{N}\rbrace$. Performing the 
$\bar{\varepsilon}$ integration by contour in the complex plane (see Appendix \ref{app_integrations}) 
leads to
\begin{equation}
\label{eq:I1eame}
I_{1,\varepsilon,a}^{(1)}  = \l[{\cal I}_{1,\varepsilon}^{(1)}\r]_{a a} \ ,
\end{equation}
with the matrix ${\cal I}_{1,\varepsilon}^{(1)}$ given by
\begin{equation}
\label{eq:I1ea0}
{\cal I}_{1,\varepsilon}^{(1)} = \frac{e}{\hbar}\, \mathrm{Im}\left\lbrace 
t^{\dagger}t\ \mathcal{V}^{11}(\varepsilon,\varepsilon)
+ t^{\dagger}r'\ \mathcal{V}^{21}(\varepsilon,\varepsilon)
\right\rbrace  \, .
\end{equation}
The $N \times N$ matrix $\mathcal{V}^{\bl l}
(\bar{\varepsilon},\varepsilon)$, spanning the space of incoming modes $\bar{l},l$, has its element 
$(\bar{a},a)$ defined by Eq.~\eqref{eq:meVT}. This definition is based on the one-to-one correspondence 
between incoming modes and outgoing scattering states. Using the unitarity of $S$ and the fact that 
$\mathcal{V}^{l,l}_{a,a}(\varepsilon,\varepsilon)$ is a real quantity we can write 
Eq.\ \eqref{eq:I1ea0} as
\begin{equation}
\label{eq:I1ea}
{\cal I}_{1,\varepsilon}^{(1)} = -\frac{e}{\hbar}\, \mathrm{Im}\left\lbrace 
r^{\dagger}r\ \mathcal{V}^{11}(\varepsilon,\varepsilon)
+ r^{\dagger}t'\ \mathcal{V}^{21}(\varepsilon,\varepsilon)
\right\rbrace  \, .
\end{equation}
The change of the total current, up to first order in $\VT$ and in linear response to the applied 
voltage $V$, is obtained [as in Eq.\ \eqref{eq:I0}] by summing the contribution of all modes at 
the Fermi energy. The first term of (\ref{eq:I1ea}) vanishes when summed over $a$ and therefore the 
${\cal O}(\VT)$ change of the zero-temperature dimensionless conductance is \cite{JSTW_2010}
\begin{equation}
\label{eq:deltag1}
g^{(1)}  = - 4 \pi\ \mathrm{Im}\left\lbrace \mathrm{Tr}
\left[r^{\dagger}t'\ \mathcal{V}^{21}\right]\right\rbrace \, ,  
\end{equation}
where all quantities are evaluated at $\varepsilon_{\rm F}$ and the trace is over the incoming, 
right-moving modes.
This result is valid for the general situation where quantum transport through a scatterer is modified 
by a weak perturbation. The matrix $r^\dagger t'$ depends only on the unperturbed scatterer, while 
the tip's effect appears in the $\mathcal{V}$ matrix elements. 

The conductance change is invariant under a change of current direction as well as under the changes 
$\VT(x,y)\to \VT(\pm x,\pm y)$ of the perturbing potential, provided the unperturbed system verifies 
the corresponding (up-down and/or left-right) reflection symmetries.

The conductance's sensitivity to electrostatic potential variations was considered in a 
one-dimensional geometry in Ref.~\onlinecite{gasparian96}. The example of a $\delta$-function barrier 
perturbed by a local tip can be analytically calculated. We show in Appendix \ref{app_ccfd} that this 
particular case is in agreement with the general expression (\ref{eq:deltag1}).

Like $g^{(0)}$, the conductance correction $g^{(1)}$ is given by the trace over the space of 
propagating modes in one lead. The obvious requirement of a base-independent result is thus fulfilled. 
We stress that, even if the matrix $\mathcal{V}^{21}$ is obtained from the overlaps of scattering states, 
the entries of its matrix elements are mode indices. The relationship between scattering states 
and lead modes is fixed by Eq.\ \eqref{allscats}, and a base transformation in the space of the modes of 
lead 1 (2) induces a change of $\Psi_{1,\varepsilon,a}$ and $\Psi_{2,\varepsilon,a}$ for $x<0$ ($x>0$).

Using the polar decomposition \eqref{eq:Spol} of the scattering matrix we can write
\begin{equation}
\label{eq:deltag1eigen}
g^{(1)}  = 4 \pi\ \sum_{m=1}^{N} {\cal R}_m {\cal T}_m \ 
\mathrm{Im}\left\lbrace \mathcal{U}^{21}_{mm} \right\rbrace \, .  
\end{equation}
Where $\mathcal{U}^{\bl l} (\bar{\varepsilon},\varepsilon)=\bar{u}_{\bl}
\mathcal{V}^{\bl l} (\bar{\varepsilon},\varepsilon)u_{l}^{\dagger}$ 
represents the matrix $\mathcal{V}^{\bl l}$ upon transformation into the 
transmission eigenmodes basis \eqref{allleadsteigen}. In Eq.\ \eqref{eq:deltag1eigen}, as in 
Eq.\ \eqref{eq:deltag1}, the energy arguments are understood to be taken at $\EF$.

From Eq.\ \eqref{eq:deltag1eigen} we see that the $m$\textsuperscript{th} mode contribution to the 
conductance correction $g^{(1)}$ is appreciable only if $m$ is partially opened. This observation 
is of foremost importance in the case of QPCs, where the transmission eigenmodes open one by one. 
On the $m$\textsuperscript{th} conductance plateau we have ${\cal T}_m \simeq 1$ and 
${\cal R}_m \simeq 0$, therefore $g^{(1)}$ is suppressed. The first order conductance correction 
is only relevant in the vicinity of a conductance step, and only the contribution arising from the 
transmission eigenmode that is partially open matters. In order to capture the dominant correction 
on the conductance plateaus, it is necessary to go beyond the first order Born approximation. 

\section{Second-order perturbation in the tip potential}
\label{sec_2nd}

The second order correction in $\VT$ to the scattering state of 
Eq.~(\ref{eq:Lippmann-Schwinger}) is obtained as
\begin{widetext}
\begin{equation}
\label{eq:second_order}
\Psi_{l,\varepsilon,a}^{(2)}(\br)=
\int \dif \bar{\br} \int \dif \barbarbr \ 
\mathcal{G}^{(0)}(\br,\bar{\br},\varepsilon) \ \VT(\bar{\br}) \
\mathcal{G}^{(0)}(\bar{\br},\barbarbr,\varepsilon) \ 
\VT(\barbarbr) \ \Psi_{l,\varepsilon,a}^{(0)}(\barbarbr) \ .
\end{equation}
The second-order change of the current density associated with the scattering state $l,\varepsilon,a$ is
\begin{equation}
\label{eq:j1bis}
j_{l,\varepsilon,a}^{x(2)}(\br) = 
j_{l,\varepsilon,a}^{x(2)\alpha}(\br)+
j_{l,\varepsilon,a}^{x(2)\beta}(\br) 
\end{equation}
with
\begin{subequations}
\label{eq:j2}
\begin{eqnarray}
\label{eq:j2a}
j_{l,\varepsilon,a}^{x(2)\alpha}(\br) &=& 
\frac{e\hbar}{\Me} \ 
\mathrm{Im}\left\lbrace 
\Psi_{l,\varepsilon,a}^{(0)*}(\br) \frac{\partial}{\partial x}
\Psi_{l,\varepsilon,a}^{(2)}(\br) -
\Psi_{l,\varepsilon,a}^{(2)}(\br) 
\frac{\partial}{\partial x} \Psi_{1,\varepsilon,a}^{(0)*}(\br) \right\rbrace \nonumber \\ &=&
2 \sum_{\bl,\bbl=1}^{2} \mathrm{Re}\left\lbrace
\int_{\tepsilon_1}^{\infty} 
\frac{\dif\bepsilon}{\varepsilon^{+}-\bepsilon} \
\int_{\tepsilon_1}^{\infty} 
\frac{\dif\bbepsilon}{\varepsilon^{+}-\bbepsilon} \ 
\sum_{\ba=1}^{\bar{N}} \sum_{\bba=1}^{\bar {\bar{N}}} 
\left[j^{x(0)}(\bar{\br})\right]_{a \ba}^{l \bl}(\varepsilon,\bar{\varepsilon})\
\left[\VT\right]^{\bl \bbl}_{\ba \bba}(\bepsilon,\bbepsilon) 
\left[\VT\right]^{\bbl  \bl}_{\bba \ba}(\bbepsilon,\bepsilon) 
\right\rbrace ,
\\
\label{eq:j2b}
j_{l,\varepsilon,a}^{x(2)\beta}(\br) &=& 
\frac{e\hbar}{\Me} \ 
\mathrm{Im}\left\lbrace \Psi_{l,\varepsilon,a}^{(1)*}(\br) \frac{\partial}{\partial x}
\Psi_{l,\varepsilon,a}^{(1)}(\br)
\right\rbrace \nonumber \\ &=&
\frac{e\hbar}{\Me}  
\sum_{\bl,\bbl=1}^{2} \mathrm{Im}\left\lbrace
\int_{\tepsilon_1}^{\infty} 
\frac{\dif\bepsilon}{\varepsilon^{-}-\bepsilon} 
\int_{\tepsilon_1}^{\infty} 
\frac{\dif\bbepsilon}{\varepsilon^{+}-\bbepsilon} 
\sum_{\ba=1}^{\bar{N}} \sum_{\bba=1}^{\bar {\bar{N}}} 
\left[\VT\right]^{l \bl}_{a \ba}(\varepsilon,\bbepsilon) 
\Psi_{\bl,\bepsilon,\ba}^{(0)*}(\br) \frac{\partial}{\partial x}
\Psi_{\bbl,\bbepsilon,\bba}^{(0)}(\br) 
\left[\VT\right]^{\bbl l}_{\bba a}(\bbepsilon,\varepsilon)
\right\rbrace .
\end{eqnarray}
\end{subequations}
The calculation of the corresponding current corrections follows the lines presented in the last section,
but it is considerably more involved due to the double energy integrations in Eq.\ \eqref{eq:j2}. 
Details are provided in Appendix \ref{app_secondorder}.

The second-order correction in $\VT$ to the total current in linear response to the applied bias $V$ 
is obtained by summing $I_{1,\varepsilon,a}^{(2)\alpha}$ and $I_{1,\varepsilon,a}^{(2)\beta}$ over all 
scattering states at the Fermi energy. Since the matrix 
$t^\dagger t\mathcal{V}^{1 \bbl}(\varepsilon,\bbepsilon)\mathcal{V}^{\bbl 1}(\bbepsilon,\varepsilon)$
has a real trace, the second-order change of the zero-temperature dimensionless conductance 
reads \cite{corr1}
\begin{eqnarray}
g^{(2)}  & = & - 4 \pi^2\ \left(  
\mathrm{Tr}\left[t^{\dagger}t\ \mathcal{V}^{12}\mathcal{V}^{21}
- r'^{\dagger}r'\ \mathcal{V}^{21}\mathcal{V}^{12}\right] -
2 \mathrm{Re}\left\lbrace \mathrm{Tr}\left[t^{\dagger}r' \mathcal{V}^{21}\mathcal{V}^{11}
\right] \right\rbrace \right)
\nonumber \\
&-& 4 \pi \sum_{\bl=1}^{2} \mathrm{Im}\left\lbrace
\int_{\tepsilon_1}^{\infty} 
\frac{\dif\bepsilon}{\bepsilon-\EF^{+}} \
\mathrm{Tr}[
t^{\dagger}r' \mathcal{V}^{2 \bl}(\EF,\bepsilon)
\mathcal{V}^{\bl 1}(\bepsilon,\EF)]
\right\rbrace
\, . 
\label{eq:deltag2alt}
\end{eqnarray}
The first three contributions, where the energy dependence is not explicit, are on-shell terms at $\EF$, while the last contribution involves an energy integration characteristic of second-order perturbation theory.

Separating in the energy integral the principal-part and delta-function contributions, we can write
\begin{equation}
\label{eq:deltag2}
\begin{aligned}
g^{(2)} = & - 4 \pi^2\ \left(  
\mathrm{Tr}\left[t^{\dagger}t\ \mathcal{V}^{12}\mathcal{V}^{21}
- r'^{\dagger}r'\ \mathcal{V}^{21}\mathcal{V}^{12}\right] +
\mathrm{Re}\left\lbrace \mathrm{Tr}\left[t^{\dagger}r'
\left(\mathcal{V}^{22}\mathcal{V}^{21}
-\mathcal{V}^{21}\mathcal{V}^{11}\right) \right] \right\rbrace \right) 
\\
& - 4 \pi \sum_{\bl=1}^{2}
{\cal P} \int_{\tepsilon_1}^{\infty} 
\frac{\dif\bepsilon}{\bepsilon-\EF} \
\mathrm{Im}\left\lbrace \mathrm{Tr}[
t^{\dagger}r' \mathcal{V}^{2 \bl}(\EF,\bepsilon)
\mathcal{V}^{\bl 1}(\bepsilon,\EF)]
\right\rbrace
\, . 
\end{aligned}
\end{equation}
\end{widetext}
As in the case of the first order correction $g^{(1)}$, $g^{(2)}$ is given 
by a trace over the incoming right-moving modes, and thus independent of the basis.
In the transmission eigenmodes basis the computation of the trace simplifies: the terms 
$r^{\dagger}t'$ are relevant outside the regime of conductance quantization, but in such a case the 
contribution $g^{(2)}$ is negligible (at small $\VT$) with respect to $g^{(1)}$. Thus, $g^{(2)}$ 
dominates only on the conductance plateaus of a QPC. In this case Eq.~(\ref{eq:deltag2}) 
is considerably simplified and becomes
\begin{eqnarray}
g^{(2)}  &=& - 4 \pi^2\   
\mathrm{Tr}\left[t^{\dagger}t\ \mathcal{V}^{12}\mathcal{V}^{21}\right]\nonumber\\
&=& - 4 \pi^2\ \sum_{m=1}^{M} \left[\mathcal{U}^{12}\mathcal{U}^{21} \right]_{m m} \, , 
\label{eq:deltag2plateau}
\end{eqnarray}
the sum running over the $M$ open eigenchannels of the QPC.

Equations \eqref{eq:deltag1} and \eqref{eq:deltag2} provide all the information about the 
effect of a non-invasive SGM tip applicable to situations without and with conductance quantization, 
respectively. While some general conclusions can be directly extracted from these expressions, 
their dependence on the unknown matrix elements $\mathcal{V}^{\bl,l}$ makes it difficult 
to perform explicit calculations and to address important questions like the locality of the response. 
In the following section we consider the particular case of an abrupt QPC since in this system the 
scattering states can be calculated explicitly and thus the evaluation of the matrix elements becomes 
possible.  

\section{Scattering states of an abrupt quantum point contact}
\label{sec_QPC}

QPCs constitute a paradigm of quantum transport since they lead to the interesting phenomenon of 
conductance quantization.\cite{vanwees88,wharam88,lindelof08} As a consequence, they have been 
intensively studied with the SGM 
technique.\cite{topinka00a,topinka01a,aidala07,jura07,jura09a,jura10,schnez11b,aoki12,kozikov12}
\begin{figure}
\includegraphics[width=.45\textwidth]{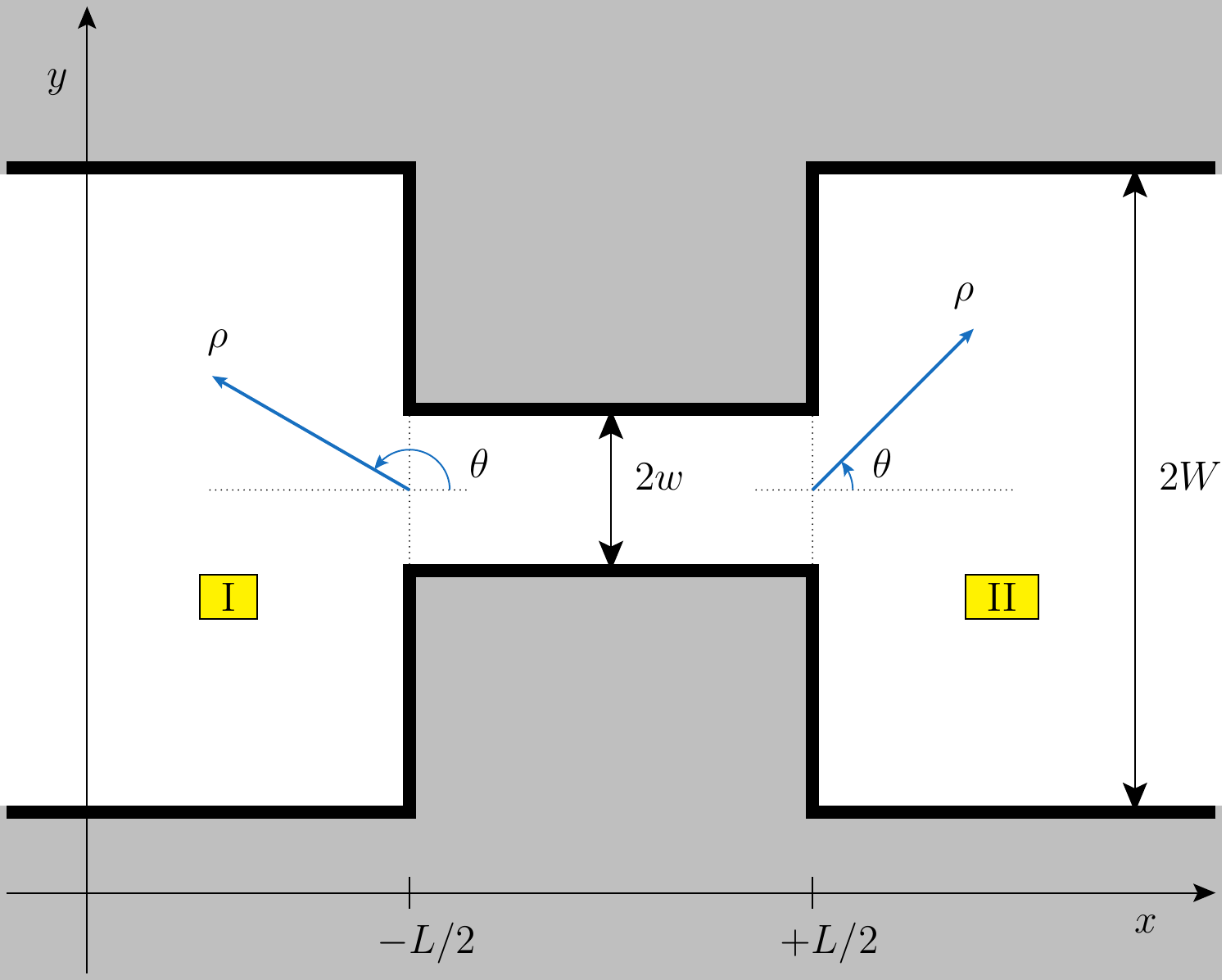}
\caption{(Color online) The wide-narrow-wide geometry representing an abrupt QPC. 
The 2DEG (white area) is delimited by gates modeled as hard-wall boundaries (thick black lines). 
The two sketched systems of polar coordinates $\rho, \theta$ are used in Sec.~\ref{sec_TEQPC}
to express the scattering eigenstates in regions I and II.}
\label{fig_setup}
\end{figure}
In order to quantify the conductance corrections of Eqs.\ \eqref{eq:deltag1} and 
\eqref{eq:deltag2}, we need the scattering states of the unperturbed system. 
While the conductance quantization of a QPC is very robust with respect to the details of its 
geometry,\cite{vanwees88,wharam88,lindelof08,glazman,szafer89,kirczenow89,buttiker90}
the scattering states are highly dependent on the details of the constriction. In this section we focus 
on the particular case of an abrupt quantum point contact (AQPC) with a wide-narrow-wide geometry as 
shown in 
Fig.~\ref{fig_setup}. In such a setup, the conductance was calculated analytically in an approximate 
way,\cite{szafer89} giving an understanding of conductance quantization's key ingredients. We 
extend these calculations in order to obtain the scattering states, and then the matrix elements 
$\mathcal{V}^{\bl,l}$. 

The right-moving scattering states in region I and II of an AQPC (see Fig.~\ref{fig_setup}) 
have the asymptotic form given in Eq.~(\ref{scatst1}), while within the constriction, i.e. 
for $-L/2<x<L/2$, they are
\begin{equation}
\label{state_n_narrow}
\Psi_{1,\varepsilon,a}^{(0)}(\br) = 
\sum_{n=1}^{\infty}\frac{c}{\sqrt{|K_{n}|}}
\left(\gamma^+_{na}\, e^{i K_{n} x} + \gamma^-_{na}\, e^{-i K_{n} x}\right) \Phi_{n}(y)
\end{equation}
with the transverse channel wave-functions in the narrow region
\begin{equation}
\Phi_{n}(y) = \frac{1}{\sqrt{w}} \sin{\left[Q_{n}(y-w)\right]}.
\end{equation}
In analogy with Eq.\ \eqref{eq:transverse_energy_lead} the transverse energy in the narrow region 
reads $\epsilon_n^{(\mathrm{t})}=\hbar^2 Q_n^2/2\Me$ with the transverse momentum $Q_n=\pi n/2w$. 
The longitudinal momentum $K_n=\sqrt{k^2-Q_n^2}$ is real for the propagating channels and pure 
imaginary for the closed channels. In the wave-function matching at $x=\pm L/2$ the fundamental 
quantities are the overlaps between the $a$ and the $n$ transverse channel wave-functions
\begin{equation}\label{coola}
\coola_{na} = \int_{-w}^{w}\dif y\,\Phi_{n}(y)\phi_a(y)
= \frac{2 Q_n}{\sqrt{wW}[q_a^2-Q^2_n]}f_n(q_a w)
\end{equation}
where
\begin{equation}\label{eq:fndef}
f_n(z)=\left\{
\begin{array}{rl}
-\cos(z) & \mathrm{for}\ n\ \mathrm{and}\ a\ \mathrm{odd}\\
\sin(z)  & \mathrm{for}\ n\ \mathrm{and}\ a\ \mathrm{even}
\end{array}
\right..
\end{equation}
By symmetry there is no coupling of channels with different parity. Notice that the $\coola_{na}$'s 
are not singular at $q_a\simeq Q_n$, and since $\phi_a$ and $\Phi_n$ are orthonormal bases in their 
respective $y$-intervals, we have the completeness relation 
$\sum_{a}\coola_{na}\coola_{n'a}=\delta_{nn'}$. 

There are two important observations at this point: (i) the $\coola_{na}$'s are appreciably 
different from zero only for $q_a\in\Delta Q_n\equiv[Q_{n-1},Q_{n+1}]$ and (ii) $k_a$ is a 
smooth function of $q_a$. Based on these observations, Szafer and Stone introduced the highly 
successful mean field approximation\cite{szafer89} (MFA), where 
Eqs.~(\ref{coola})-(\ref{eq:fndef}) are simplified by taking  
\begin{equation}
\label{MFA1}
\left.\coola_{na}\right|_\mathrm{MFA}
= \left\{\begin{array}{cc}
\sqrt{\frac{w}{2W}}[1+(-1)^{n+a}] &  \mathrm{if}\ q_a\in\Delta Q_n \\
0                                 &  \mathrm{if}\ q_a\notin\Delta Q_n
\end{array} . \right.
\end{equation}
Such an approximation allows for an analytical determination of the conductance of an AQPC, 
but does not lead to the scattering states. Thus it is necessary to have a less restrictive 
approximation, a smooth field approximation (SFA) where the $q_a$ dependence in $\coola_{na}$ 
of Eq.~(\ref{coola}) is kept. 

Next, introduce a generalized momentum-like quantity crucial to 
the wave-function matching problem
\begin{equation}
%\label{Anm_dscr}
\label{Anm_cont}
{\cal K}_{nn'} \equiv {\sum_a}'k_a \coola_{na}\coola_{n'a}.
\end{equation}
The prime over the summation symbol indicates that only the $a$'s with the same parity as $n$ 
are considered. When going to the continuum limit in the wide region the density of modes $q$ 
is $W/\pi$ -- i.e. only half the usual one. This leads to
\begin{equation}
{\cal K}_{nn'} = \frac{W}{\pi}\int_0^\infty \dif q\, k^{(\mathrm{l})}(q) \coola_{n}(q)\coola_{n'}(q) ,
\end{equation}
where $k^{(\mathrm{l})}(q)=\sqrt{k^2-q^2}$ is the longitudinal wave-vector. 
Taking into account the smoothness condition (ii) and the completeness relation of the 
$\coola_{na}$'s gives
\begin{equation}
\label{Anm_approx}
{\cal K}_{nn'} \approx {\cal K}_n \sum_{a}\coola_{na}\coola_{n'a} = {\cal K}_n\delta_{nn'}.
\end{equation}
Consistently with the above mentioned property (i), we take ${\cal K}_n$ as the average value 
of the longitudinal wave-vector in the interval $\Delta Q_n$, that is
\begin{equation}
{\cal K}_n = \frac{w}{\pi}\int_{\Delta Q_n}\dif q\,k^{(\mathrm{l})}(q).   
\end{equation}
According to the positioning of $k$ with respect to the interval $\Delta Q_n$, the generalized 
wave-vector ${\cal K}_n$ may have real and/or imaginary parts. 

The diagonal form (\ref{Anm_approx}) greatly simplifies the wave-function matching problem, 
leading to
\begin{eqnarray}
\label{eq:gamma_na}
\gamma^\pm_{na} &=& \frac{2 \sqrt{k_a K_n}(K_n\pm {\cal K}_n)\coola_{na}}{D_n} \, e^{-i(k_a\pm K_n)L/2},
\\
\label{eq:D_n}
D_{n} &=& (K_n+{\cal K}_n)^2\, e^{-iK_nL} - (K_n-{\cal K}_n)^2 \, e^{iK_nL} .
\end{eqnarray}
The transmission and reflection amplitudes, respectively, are
\begin{subequations}
\begin{eqnarray}
\label{t_matrix}
t_{ba} &=& 4\sqrt{k_b k_a}\,e^{-i\l[(k_{b}+k_a)L/2\r]} \sum_n \frac{K_n}{D_n}\coola_{nb}\coola_{na},
\\ \nonumber
r_{ba} &=& -\delta_{ba}e^{-ik_bL}  \\
\label{r_matrix}
&&+ 2\sqrt{k_b k_a}e^{-i\l[(k_b+k_a)L/2\r]} \sum_n \frac{B_n}{D_n}\coola_{nb}\coola_{na}
\end{eqnarray}
\end{subequations}
with
\begin{equation}\label{eq:bn}
 B_n=(K_n+{\cal K}_n)e^{-iK_nL}+(K_n-{\cal K}_n)e^{iK_nL}.
\end{equation}
The dimensionless conductance naturally acquires the form \eqref{eq:geigen} with
\begin{equation}
\label{SSeigen}
{\cal T}_m = \frac{4 |K_m|\, {\rm Re}\lbrace {\cal K}_m\rbrace}{|D_m|}\, ,
\end{equation}
explicitly showing the one-to-one correspondence between the transmission eigenmodes $m$ and the 
constriction channels $n$. Therefore, from now on we will make the identification $n=m$. Notice 
that Eq.~\eqref{SSeigen} takes into account the contribution of the $M$ open ($Q_n<k$) and the 
$N-M$ evanescent ($Q_n>k$) channels of the QPC, since for the lowest evanescent mode 
${\rm Re}\lbrace {\cal K}_n \rbrace \neq 0$. 

\begin{figure}
\includegraphics[width=.45\textwidth]{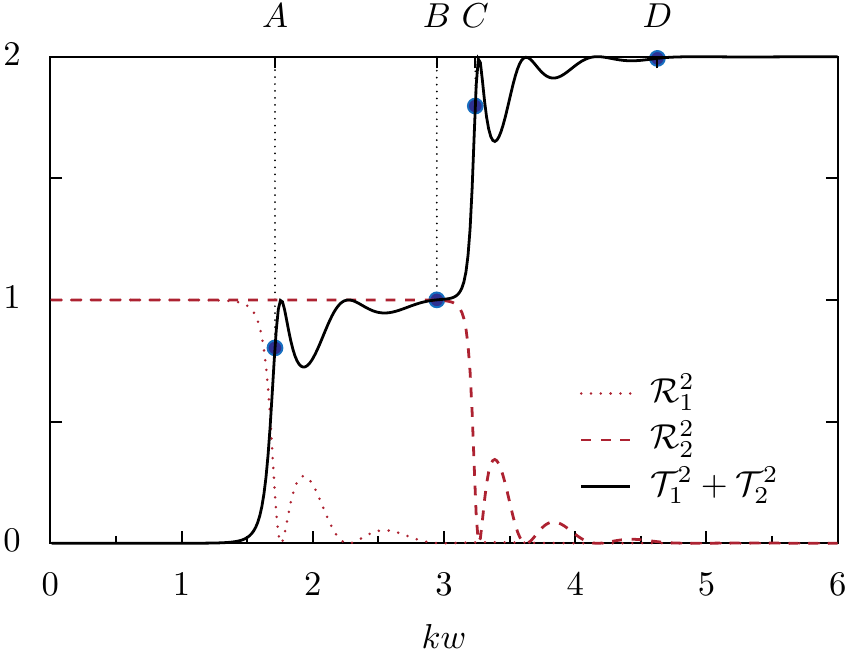}
\caption{(Color online) Total transmission (solid) and reflection eigenvalues of the first (dotted) and second (dashed) 
eigenmodes of an AQPC, as a function of the wave-vector $k$, 
for the lowest 2 channels [obtained from Eq.~\eqref{SSeigen} of the SFA]. 
The blue circles indicate the $k$-values
used in Figs.~\ref{fig_magicmode1}-\ref{fig_dg_plateaus}.}
\label{fig_conductance}
\end{figure}

The conductance resulting from Eq.\ \eqref{SSeigen} is shown in Fig.~\ref{fig_conductance} and 
agrees with that obtained under the more restrictive assumptions of the MFA, providing an 
extremely good approximation for the numerically calculated conductance of an AQPC.\cite{szafer89} 
The conductance oscillations as a function of $k$ result from quantum interference occurring 
in the AQPC.\cite{kirczenow89} A finite temperature provides considerable smoothing and an 
improvement of the conductance quantization.\cite{lindelof08}

The scattering states are completely determined by Eqs.\ \eqref{eq:gamma_na}-\eqref{eq:bn},  
putting us in a position to readily evaluate the matrix elements \eqref{eq:meVT} once the tip 
potential is specified.
In fact, as we saw in Secs.\ \ref{sec_1st} and \ref{sec_2nd}, the expressions for 
$g^{(1)}$ and $g^{(2)}$ simplify considerably in the transmission eigenmodes basis,
which will then be our choice in the next section.

\section{Scattering eigenstates of an abrupt quantum point contact}
\label{sec_TEQPC}

In the same way as the incoming modes \eqref{allleadst} generate the outgoing scattering states 
\eqref{allscats}, the transmission eigenmodes \eqref{allleadsteigen} give rise to 
{\it scattering eigenstates} $\chi_{l,\varepsilon, m}$ that are eigenfunctions of the current 
operator ${\cal I}_{l,\varepsilon}^{(0)}$. Determining
$\varrho_{l,\varepsilon,m}^{(-)}$ requires the diagonalization of $t^{\dagger} t$ and
$t^{\prime \dagger} t^{\prime}$, which is, in general, a difficult task. The knowledge of 
the scattering states of an AQPC developed in the last section will be used to derive 
approximate expressions for the  scattering eigenstates. 
The latter can be written as linear combinations of the scattering states
\begin{equation}
\label{psi_magic1}
\chi_{l,\varepsilon, m}(\br)=\sum_a\, c^{(m)}_{l,\varepsilon,a}\,\Psi^{(0)}_{l,\varepsilon,a}(\br).
\end{equation}
The coefficient $c^{(m)}_{1(2),\varepsilon,a}$ coincides with the matrix element 
$\left[u_{1(2)}\right]_{ma}^{*}$ of Eq.\ \eqref{allleadsteigen} up to an overall $m$-dependent phase.

From the form \eqref{allscats} of the scattering wave-functions and the form \eqref{t_matrix} 
of the transmission amplitudes, the general expression for the scattering eigenstates in region II is
\begin{equation}
\label{building_magic} 
\chi^{\mathrm{II}}_{1,\varepsilon, m}(\br) = c
\sum_{n}\,\frac{4K_n}{D_n} \, A_{mn} F_m(\br)\, .
\end{equation}
The spatial dependence of $\chi^{\mathrm{II}}_{1,\varepsilon, m}$ is through the function
\begin{equation}
F_m(\br)=\sum_{b}\coola_{n,b} \ \phi_{b}(y) \ e^{i k_{b} (x-L/2)}\, , 
\end{equation}
while the prefactor $A_{mn}$ is defined by
\begin{equation}
A_{mn}=\sum_a c^{(m)}_{1,\varepsilon,a}\sqrt{k_a} \ \coola_{n a} \ e^{-i k_a L/2}\, .
\end{equation}
The form 
\begin{equation}
\label{cm}
c^{(m)}_{1,\varepsilon,a}=\sqrt{\frac{k_a}{{\rm Re}\lbrace{\cal K}_m\rbrace}} \ \coola_{ma} \ e^{ik_aL/2} 
\end{equation}
satisfies the normalization condition $\sum_a \left|c^{(m)}_{1,\varepsilon,a}\right|^2=1$ and leads to
\begin{equation}\label{norm_magic}
A_{mn}=\sqrt{\mathrm{Re}\left\{\mathcal{K}_{m}\right\}} \ \delta_{mn}\, ,
\end{equation}
ensuring that $\chi^{\mathrm{II}}_{1,\varepsilon, m}$ is an eigenvector of $\mathcal{I}_1^{(0)}$
with an associated current eigenvalue
\begin{equation}
\l[{\cal I}^{(0)}_1\r]_{mm}=\frac{e}{h} \ \frac{(4|K_m|)^2|A_{mm}|^2
\, \mathrm{Re}\lbrace{\cal K}_{m}\rbrace}{|D_m|^2} \ .
\end{equation}
Since $\l[{\cal I}^{(0)}_{1,\epsilon}\r]_{mm}=\frac{e}{h}{\cal T}_{m}^2$ the choice 
(\ref{cm}) for the coefficients $c^{(m)}_{1,\varepsilon,a}$ is consistent with the form (\ref{SSeigen}) 
of the transmission eigenvalues. The $c^{(m)}_{1,\varepsilon,a}$ are obtained up to 
an overall, $m$-dependent phase, as it is usually the case in the determination of eigenvectors. 
Given the simple form (\ref{norm_magic}) of $A_{mn}$, 
the wave-function (\ref{building_magic}) of the right-moving scattering eigenstate reads 
\begin{equation}
\label{psi_magic_II}
\chi^{\mathrm{II}}_{1,\varepsilon, m}(\br) = 
\frac{c}{\sqrt{{\rm Re}\lbrace{\cal K}_m\rbrace}} \
t_m \ F_m(\br),
\end{equation}
having defined $t_m=(4K_m{\rm Re}\lbrace{\cal K}_m\rbrace)/D_m$. 
In general $|t_m|={\cal T}_{m}$, while for the propagating channels $t_m={\cal T}_{m}$.

$F_m(\br)$ can be calculated by going to the continuum limit in the sum over the transverse channels 
\begin{equation}
\label{stat_phase_1}
F_m(\br) = \frac{2(-1)^m}{\pi w^{1/2}}\int_0^{k}\!\dif q\,
f_m(qy)\,e^{ik^{(\mathrm{l})}(q)(x-L/2)}\,
\frac{Q_m f_m(qw)}{q^2-Q_m^2} .
\end{equation}
Since the function $f_m$ defined in Eq.\ \eqref{eq:fndef} is trigonometric we can split
$F_m(\br)$ in two integrals and write
\begin{equation}
F_m(\br) = \varsigma_m [F^{+}_m(\br)+ 
(-1)^{m-1}F^{-}_m(\br)] \ ,
\end{equation}
with $\varsigma_m = 1$ ($-i$) for odd (even) $m$ and
\begin{equation}
F^\pm_m(\br)=\frac{1}{\pi w^{1/2}} \int_0^{k} \dif q\, g_m(q)\, e^{ih_{\pm}(q)} .
\end{equation}
The function
\begin{equation}
g_m(q)=\frac{Q_m}{q^2-Q_m^2} \ f_m(qw)
\end{equation}
has a smooth dependence on $q$ (even for $q\simeq Q_m$), whereas $h_{\pm}$ is defined as
\begin{equation}
h_\pm(q)=k^{(\mathrm{l})}(q)(x-L/2)\pm qy.
\end{equation}
Using two-dimensional polar coordinates in region II (I) with origin at the center of the 
right (left) opening one can write 
${\boldsymbol{\rho}} = {\bf r} \mp (L/2){\bf \hat {x}} = \rho\left(\cos\theta,\sin\theta\right)$, where 
$\theta~\in~[0,\pi/2] \cup [3\pi/2,2\pi]$ for II and $\theta~\in~[\pi/2,3\pi/2]$ for I. 
Expressing the wave-vector in polar coordinates
$(k^{(\mathrm{l})}(q),q)=k(\cos\vartheta,\sin\vartheta)$ one can write
\begin{equation} h_\pm(q)= k\rho\l[\cos\vartheta\cos\theta
\pm\sin\vartheta\sin\theta\r].
\end{equation}
For $k \rho\gg1$ the functions $h_\pm$ are rapidly varying with $q$ (or $\vartheta$), and therefore 
$F^{\pm}_m(\br)$ can be computed in the stationary phase approximation leading to
\begin{equation}\label{eq:PsimagicII}
\chi^\mathrm{II}_{1,\varepsilon,m} (\br)
= c \ \frac{\varsigma_m e^{-i\pi/4}}{\sqrt{{\rm Re}\lbrace{\cal K}_m\rbrace}} 
\sqrt{\frac{2}{\pi w}} \
t_m \ \frac{e^{ik \rho}}{\sqrt{k \rho}} \
\Theta_m(k,\theta) \ ,
\end{equation}
with
\begin{equation}\label{eq:defTheta}
\Theta_m(k,\theta) = \frac{k Q_m\cos\theta}
{(k\sin\theta)^2-Q_m^2} \ f_m(kw\sin\theta) \ .
\end{equation}
Notice that the condition $k\rho\gg1$ means that our results hold already at distances from 
the opening of the order of a Fermi wave-length. The asymptotic form of the scattering 
eigenstates has the form of a radial outgoing wave $e^{ik \rho}/\sqrt{k \rho}$, with an 
angular modulation $\Theta_m(k,\theta)$ peaked along the directions $\theta$ for which 
there is a mode matching between the transverse wave-vectors $Q_m$ and $k |\sin{\theta}|$.

Using the coefficients $c^{(m)}_{1,\varepsilon,a}$ of Eq.~(\ref{cm}), a similar 
stationary-phase integration over the transverse channels $q$ yields the form of 
$\chi_{1,\varepsilon, m}$ in region I
\begin{widetext}
\begin{equation}\label{eq:PsimagicI}
\chi^\mathrm{I}_{1,\varepsilon, m} (\br) = -c \ 
\frac{\varsigma_m e^{i\pi/4}}{\sqrt{{\rm Re}\lbrace{\cal K}_m\rbrace}} \
 \  \sqrt{\frac{2}{\pi w}} \frac{1}{\sqrt{k \rho}}
\left( e^{-ik \rho}+r_m e^{ik \rho} \right) \
\Theta_m(k,\theta) \ .
\end{equation}
\end{widetext}
The reflection amplitude $r_m$ reads
\begin{equation}
r_m=\frac{2B_m{\rm Re}\lbrace{\cal K}_m\rbrace}{D_m}-1 \ ,
\end{equation}
and is such that ${\cal R}_m^2=|r_m|^2=1-|t_m|^2$. The reflection and transmission eigenvalues 
of an AQPC are plotted in Fig.~\ref{fig_conductance}.

\begin{figure}
\includegraphics[width=\linewidth]{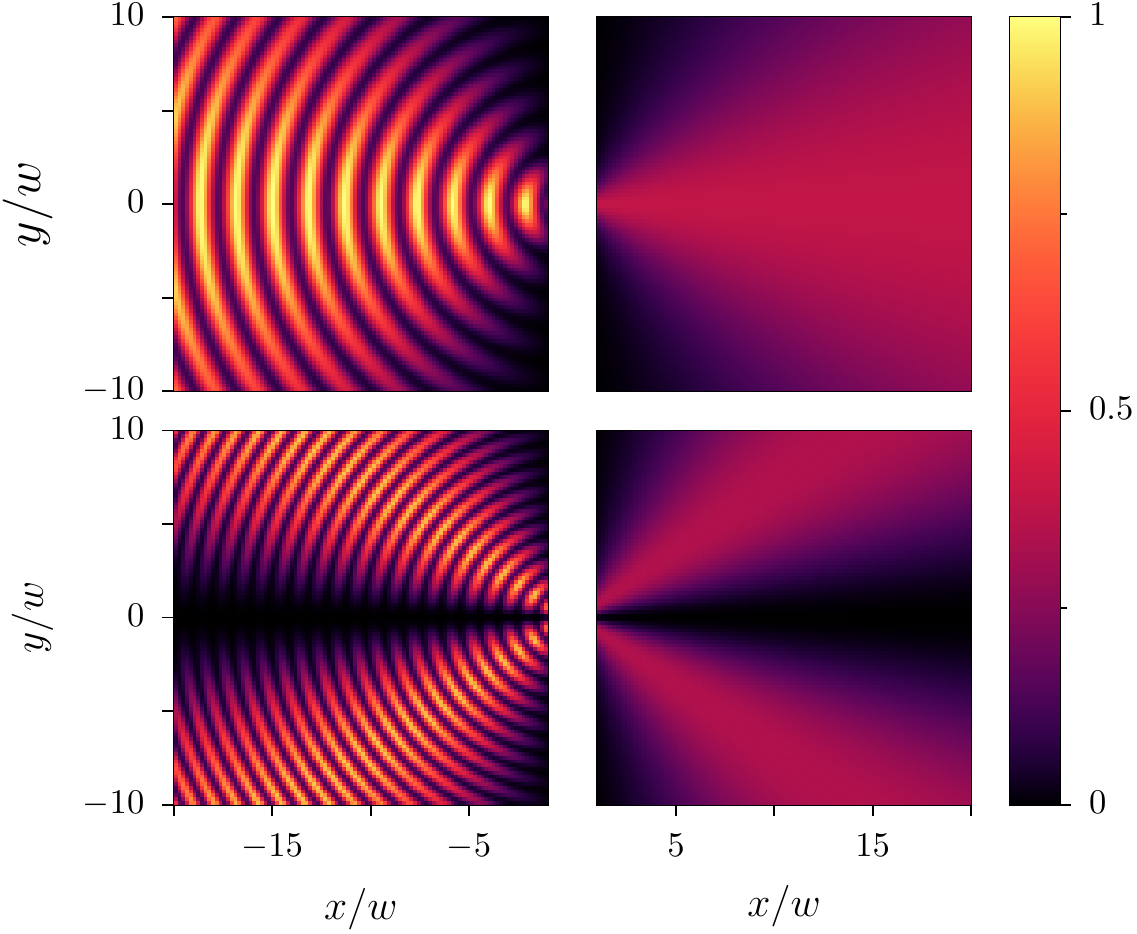}
\caption{(Color online) Color scale representing the squared absolute value of the scattering 
eigenstate wave-functions, multiplied by $\rho$, in region I [left panels, 
Eq.\ \eqref{eq:PsimagicI}] and II [right panels, Eq.\ \eqref{eq:PsimagicII}], when the energy 
$\varepsilon$ is such that the particular mode is opening (i.e.\ at the conductance steps).
The upper and lower panels, respectively, show $\rho|\chi_{1,\varepsilon,1} (\br)|^2$ and 
$\rho|\chi_{1,\varepsilon,2} (\br)|^2$, with a scale given by the extreme values of the former.
The corresponding value of $k$ considered in the upper (lower) panels is marked 
as ``$A$'' (``$C$'') in Fig.~\ref{fig_conductance} and has 
${\cal R}^2_1\approx0.2$ (${\cal R}^2_2\approx0.2$). The factor $\rho$ compensates for the 
$1/\rho$ decay of the squared modulus and is introduced in order to make the wave-function 
profile visible at large distances. The distances $x$ and $y$ are scaled with the width $w$ 
of the constriction, $L=2.5\, w$.}
\label{fig_magicmode1}
\end{figure}
\begin{figure}
\includegraphics[width=\linewidth]{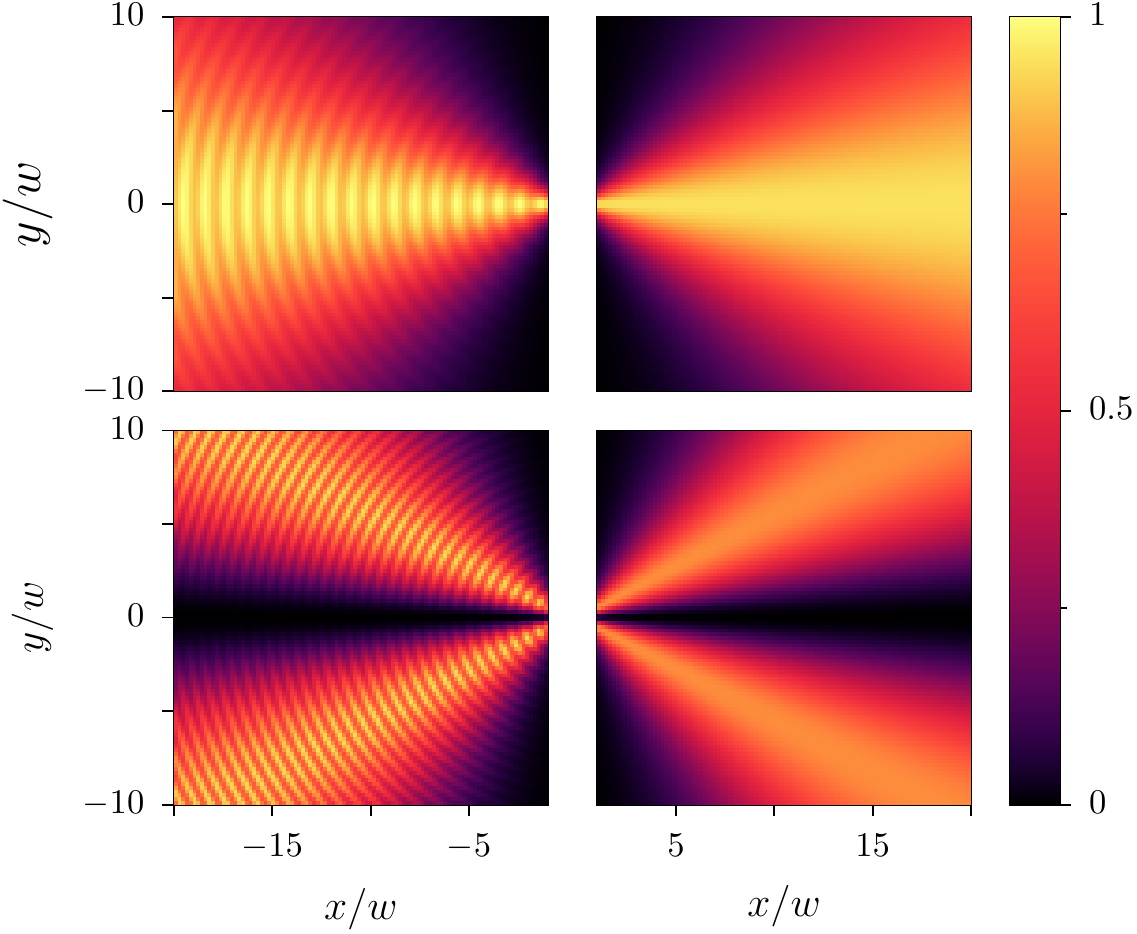}
\caption{(Color online) Same as Fig.\ \ref{fig_magicmode1}, but for energies $\varepsilon$ 
such that the particular mode is open (i.e. on the $m=1,2$ conductance plateaus).
The $k$ value for the upper (lower) panels corresponds to point ``$B$'' (``$D$'') 
from Fig.~\ref{fig_conductance}.}
\label{fig_magicmode2}
\end{figure}
Figs.~\ref{fig_magicmode1} and \ref{fig_magicmode2} show, respectively, the scattering eigenstates 
in regions I and II for values of $k$ at the step and on a plateau for $m=1$ (top panels) and $m=2$ 
(lower panels). The scattering eigenstates show all the expected features: (i) a number of lobes 
dependent on the mode index; (ii) a widening conical shape centered on the QPC (iii) a decay 
proportional to $1/\sqrt{\rho}$. At a conductance step the eigenmode is mainly reflected, 
leading to a highly asymmetric wave-function, whereas on a plateau an almost perfect 
$({\cal T}^2_m\approx1)$ transmission yields a rather symmetric pattern (with respect to the $y$-axis). 
As discussed in the previous section, the abruptness of the constriction degrades the perfect 
${\cal T}^2_m=1$ transmission \cite{szafer89,kirczenow89} 
typical of a conductance plateau, resulting in the very weak interference fringes from reflected 
waves visible in region I.

In the limit of a point-like opening, $k w\ll 2 \pi$, the problem
becomes analogous to that of two-dimensional waves diffracting at a slit,
and indeed the wave-function reduces to an angle-modulated outgoing Hankel function in the 
asymptotic limit, that is,
\begin{equation}
\chi^\mathrm{II}_{1,\varepsilon, m} (\br)
\sim \frac{e^{ik \rho}}{\sqrt{k \rho}}
\cos(\theta) \ .
\end{equation}
In the opposite classical limit of a wide constriction, $k w \gg 2 \pi$,
it instead reduces to a straight ray
\begin{equation}
\chi^\mathrm{II}_{1,\varepsilon, m} (\br)
\sim
\frac{e^{ik\rho}}{\sqrt{k\rho}}\delta(\theta) \ .
\end{equation}

Though $\chi_{1,\varepsilon, m} (\br)$ was explicitly computed separately in region I and II, 
in the case of open modes (conductance plateaus with ${\cal T}^2_m\approx1$ and 
${\cal R}^2_m\approx 0$) the central symmetry of the QPC translates into 
$\chi_{1,\varepsilon, m} (\br)=\chi^*_{1,\varepsilon, m}(-\br)$, up to the above-discussed 
residual interference fringes linked with the non-unitarity of the transmission. Thus, in a 
centrally symmetric constriction without magnetic fields, knowledge of the wave-function 
of an open eigenmode impinging from one of the leads (1 or 2) in one of the half spaces 
($x>0$ or $x<0$) is enough to evaluate the matrix elements of the perturbation. 

According to Eq.\ \eqref{eq:PsimagicII} the scattering eigenstate $\chi_{1,\varepsilon,m}$ has 
the form of a radial function times a real function containing the angular dependence. 
The corresponding current density in region II is a vector field oriented in the radial 
direction $\rho$ with absolute value
\begin{equation}\label{eq:jmagic}
j^{\rho(0),{\rm II}}_{1,\varepsilon,m}(\br)= \frac{e\hbar k}{\Me}
\left|\chi^{\rm II}_{1,\varepsilon,m}(\br)\right|^2 
\ .    
\end{equation}
The current density in region I is a radial vector field as well, whose absolute value reads
\begin{equation}\label{eq:jmagicI}
j^{\rho(0),{\rm I}}_{1,\varepsilon,m}(\br)= 
\frac{2 e}{h} 
\frac{1}{{\rm Re}\lbrace{\cal K}_m\rbrace \rho} \
\left(-1 + |r_m|^2\right) \Theta_m^2(k,\theta) \ .    
\end{equation}
On a conductance plateau $r_m=0$, and then $j^{\rho(0),{\rm I}}_{1,\varepsilon,m}(\br)$ is 
proportional to $\left|\chi^{\rm I}_{1,\varepsilon,m}(\br)\right|^2$. We stress that such a 
connection, like that of Eq.~\eqref{eq:jmagic}, only holds if: (i) the tip is at a certain 
distance from the QPC, $\rho k\gg 1$, and (ii) the system is perfectly clean. These simple 
observations have a crucial effect on the physical interpretation of SGM data, as we 
discuss in the following Section. 

\section{Conductance corrections from a local tip in the neighborhood of symmetric 
and asymmetric structures}
\label{sec_discussion}

The tip induces a local perturbing potential centered on the projection $\br_{\rm T}$ of the 
tip center on the 2DEG. For simplicity's sake consider the extreme case of a 
$\delta$-like probe
\begin{equation}
\label{eq:local_pot}
\VT(\br)=v_{\rm T}\delta(\br-\br_{\rm T}) \ .
\end{equation}
The use of more realistic smoother potential profiles would not qualitatively change 
our conclusions: their only substantial effect would be a reduction of the SGM resolution,
in the sense that the signal would be due to ``smeared'' matrix elements,
see Eq.~\eqref{eq:meVT}\cite{footnote2}.
According to Eq.\ \eqref{eq:local_pot}, 
\begin{equation}\label{eq:Vmagic}
\mathcal{U}_{{\bar m},m}^{21}(\br_{\rm })
=v_{\rm T} \ \chi^{*}_{2,\varepsilon,{\bar m}}(\br_{\rm T})
\ \chi_{1,\varepsilon,m}(\br_{\rm T}) \ .
\end{equation}
In general, there is no relationship between $\chi_{1,\varepsilon,m}$ and $\chi_{2,\varepsilon,m}$
since the matrices $u_1$ and $u_2$ of the polar decomposition \eqref{eq:Spol}, defining the 
transmission eigenmodes \eqref{allleadsteigen} and the corresponding scattering eigenstates 
\eqref{psi_magic1}, are independent. Particular cases arise in symmetric setups, where the 
scattering matrix $S$ exhibits further symmetries, beyond the condition $S^{\rm T}=S$ stemming from 
time-reversal. Our example of an AQPC is one of them, since it has the fourfold (4F) symmetry due to 
the up-down and left-right reflection symmetries. The 4F symmetry allows to write the scattering matrix 
\eqref{eq:scatt_mat} as
\begin{equation}
\label{eq:scatt_mat4F}
S = \left( \begin{array}{cc}
S_{\rm e} & 0 \\
0 & S_{\rm o}
\end{array} \right) \, ,
\end{equation}
where the $N \times N$ scattering matrix 
$S_{\rm e(o)}$ corresponding to even (odd) modes has the structure \cite{baranger96}
\begin{equation}
\label{eq:scatt_even_odd}
S_{\rm e(o)} = \left( \begin{array}{cc}
r_{\rm e(o)} & t_{\rm e(o)} \\
t_{\rm e(o)} & r_{\rm e(o)}
\end{array} \right) \, ,
\end{equation}
with $r_{\rm e(o)}^{\rm T}=r_{\rm e(o)}$ and
$t_{\rm e(o)}^{\rm T}=t_{\rm e(o)}$. In the polar decomposition of $S_{\rm e(o)}$ we see that 
$u_{1,\rm e(o)}=u_{2,\rm e(o)}$. Moreover, since 
Eqs.~\eqref{allleadst} and \eqref{eq:phi_transv} dictate that 
$\varphi_{2,\varepsilon,a}^{(-)}(x,y)=
(-1)^p \varphi_{1,\varepsilon,a}^{(-)}(-x,-y)$ for $x>0$, 
\begin{equation}
\label{eq:rel_chi1_chi2}
\chi_{2,\varepsilon,m}(\br)=(-1)^{m-1} 
\chi_{1,\varepsilon,m}(-\br) \, .
\end{equation} 
The first-order conductance correction of a 4F symmetric structure perturbed by a delta-function 
tip potential at $\br_{\rm T}$ has the form
\begin{eqnarray}
\label{eq:deltag1_4F}
g^{(1)}_{\rm 4F}(\br_{\rm T})  &=& 
4 \pi\ v_{\rm T} \sum_{m=1}^{N} (-1)^{m-1} \ 
{\cal R}_m {\cal T}_m
\nonumber\\
& & 
\ \mathrm{Im}\left\lbrace 
\chi_{1,\varepsilon,m}^{*}(-\br_{\rm T}) \
\chi_{1,\varepsilon,m}(\br_{\rm T}) \right\rbrace \, .
\end{eqnarray}

The dependence of $g^{(1)}_{\rm 4F}$ on the values of the scattering eigenstates at two points 
makes it independent of the overall $m$-dependent phase appearing in the determination of 
$\chi_{1,\varepsilon,m}$ and gives a non-local character to the conductance correction. 
However, since the two points involved are symmetric with respect to the center of the structure, 
the corresponding values are indeed related.

\begin{figure}
\includegraphics[width=\linewidth]{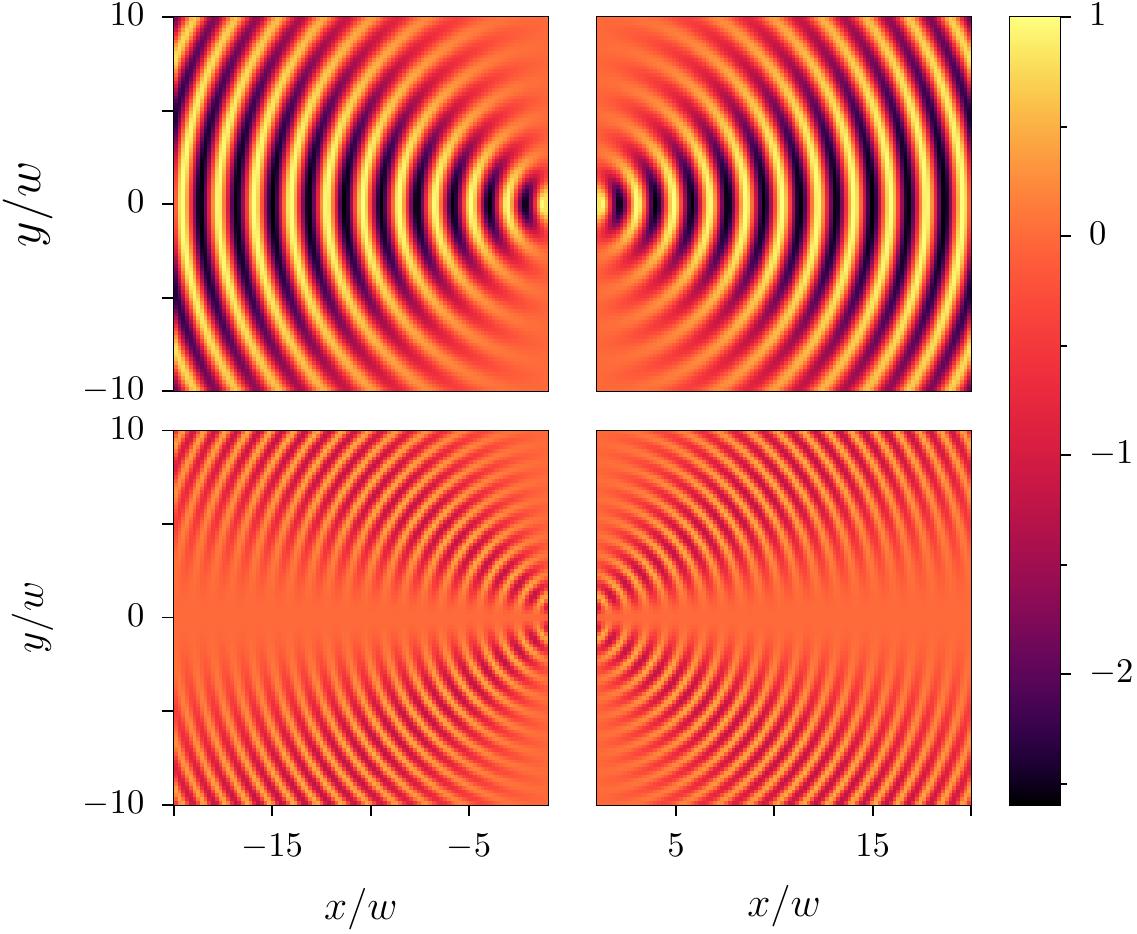}
\caption{(Color online) Conductance corrections at the lowest two steps for an AQPC from 
Eq.~\eqref{eq:deltag1_AQPC}. The $1/\rho$ decay is compensated as in 
Figs.~\ref{fig_magicmode1} and \ref{fig_magicmode2}, and the color scale is set by 
$g^{(1)}_{{\rm AQPC},1}(\br_{\rm T})$. The presence of the SGM tip at certain positions 
enhances, rather than lowering, the conductance, in agreement with the numerical simulations 
of Ref.~[\onlinecite{JSTW_2010}]. The $k$ value for the upper (lower) panels corresponds to 
point ``$A$'' (``$C$'') from Fig.~\ref{fig_conductance}.}
\label{fig_dg_steps}
\end{figure}

Our example of an abrupt QPC is a particularly simple case of a 4F structure. The condition of
Eq.\ \eqref{eq:rel_chi1_chi2} is readily checked for an AQPC since $\chi_{2,\varepsilon,m}$ can be 
obtained from  $\chi_{1,\varepsilon,m}$ by exchanging I by II and $\br$ by $-\br$ (or $\theta$ 
by $\theta+\pi$).  For an AQPC probed by a delta-function tip when the Fermi wave-vector sets the conductance at 
the $m$\textsuperscript{th} step, from Eqs.\ \eqref{eq:PsimagicII}, \eqref{eq:PsimagicI}, and 
\eqref{eq:deltag1_4F}, one obtains the first-order conductance correction 
\begin{eqnarray}\label{eq:deltag1_AQPC}
g^{(1)}_{{\rm AQPC},m}(\br_{\rm T})  &=& 
-32 v_{\rm T} \frac{c^2}{w} \ \Theta^2_m(\kF,\theta) 
{\cal R}_m{\cal T}_m
\nonumber\\
& & 
\frac{1}{\kF\rho} \mathrm{Im}\left\lbrace \frac{K_m}{D_m}
\left(e^{2i \kF \rho} + r_{m}^{*}\right) \right\rbrace,
\end{eqnarray}
which is plotted in Fig.~\ref{fig_dg_steps}.

This expression is valid in both regions I and II, and 
$g^{(1)}_{{\rm AQPC},m}(-\br_{\rm T}) = g^{(1)}_{{\rm AQPC},m}(\br_{\rm T})$. Such a result 
is consistent with the general finding of Sec.~\ref{sec_1st} that $g^{(1)}$ remains invariant 
under the changes $\VT(x,y)\to \VT(\pm x,\pm y)$ when  the unperturbed system exhibits the 
corresponding (up-down and/or left-right) symmetries.

\begin{figure}
\includegraphics[width=\linewidth]{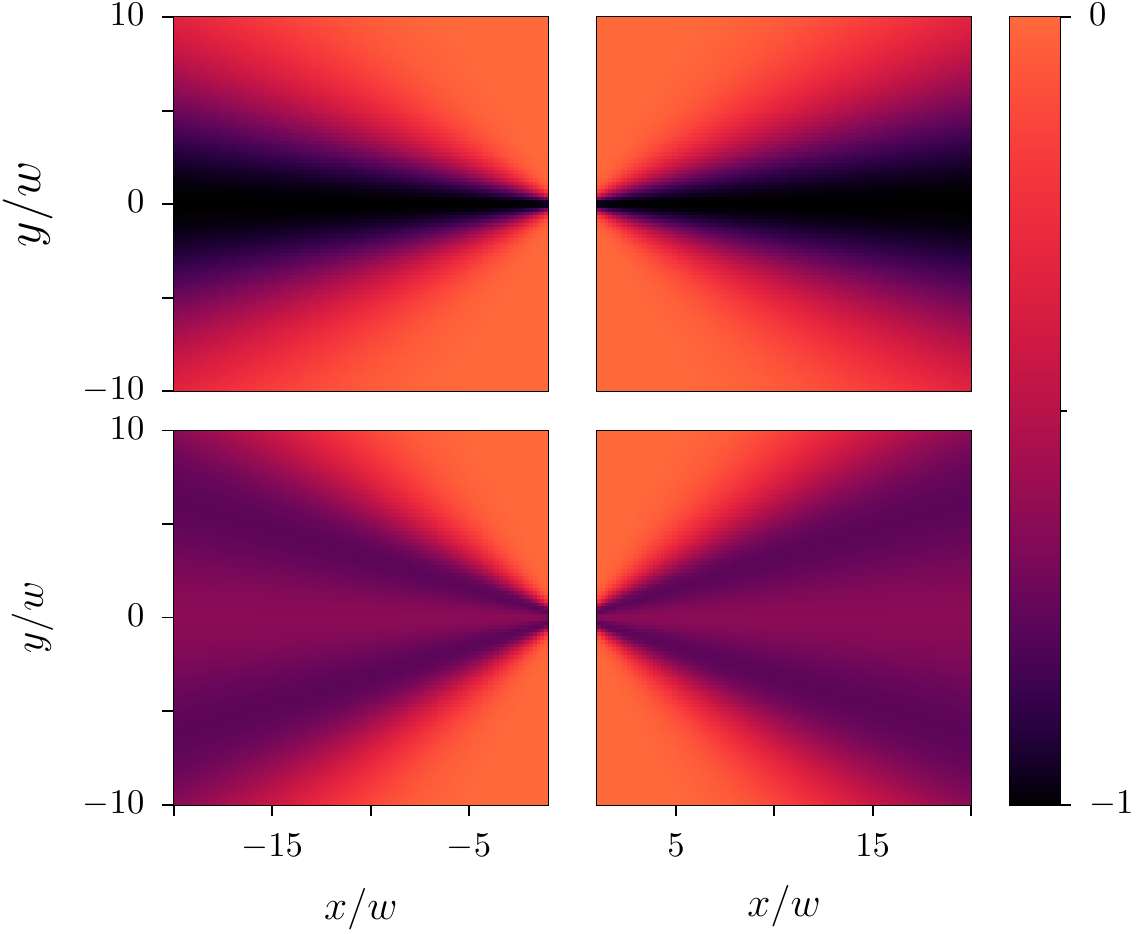}
\caption{(Color online) Same as for Fig.~\ref{fig_dg_steps}, but on the lowest two plateaus, 
Eq.~\eqref{dg_vs_current2_AQPC}. The color scale is set by $g^{(2)}_{{\rm AQPC}}(\br_{\rm T})$ 
on the first plateau. The $k$ value for the upper (lower) panels corresponds to point 
``$B$'' (``$D$'') from Fig.~\ref{fig_conductance}. The minima of the conductance correction 
correspond to the maxima of the squared absolute scattering eigenmode wave-function.}
\label{fig_dg_plateaus}
\end{figure}

The first-order conductance correction of an AQPC at the $m$\textsuperscript{th} step has the 
same angular dependence as the probability density of the corresponding scattering eigenstate, 
oscillates in the radial coordinate as $\cos{2 \kF \rho}$, and decays spatially as 
$(\kF \rho)^{-1}$. Such a behavior is in good agreement with quantum mechanical numerical 
calculations \cite{JSTW_2010}. Even if $g^{(1)}_{{\rm AQPC},m}(\br_{\rm T})$ appears as a 
local correction, it is not directly related with local observables such as the probability 
density or the current density. Moreover, we stress that the local character of 
$g^{(1)}_{{\rm AQPC},m}$ stems from the symmetry of the unperturbed structure. 
For a non-symmetric structure the first-order conductance correction is in general not local, 
and unrelated to the probability density or current density.

On a conductance plateau the dominant correction is $g^{(2)}$ and, according to 
Eq.\ \eqref{eq:deltag2plateau}, for a $\delta$-function tip at $\br_{\rm T}$
\begin{equation}
\label{dg_vs_current2}
g^{(2)}(\br_{\rm T}) = - 4 \pi^2 v_{\rm T}^2 
\sum_{m,{\bar m}=1}^{M} 
|\chi_{2,\varepsilon,{\bar m}}(\br_{\rm T})|^2 \ 
|\chi_{1,\varepsilon,m}(\br_{\rm T})|^2 \ .
\end{equation}

In a 4F symmetric QPC the condition \eqref{eq:rel_chi1_chi2}, together with the relationship 
between the probability density and the radial component of the current density \eqref{eq:jmagic}, 
yields in both regions I and II,
\begin{equation}
\label{dg_vs_current2_4F}
g_{\rm 4F-QPC}^{(2)}(\br_{\rm T}) = - 4 \pi^2 
v^2_{\rm T} \ \frac{\Me}{e\hbar \kF}
\left[\sum_{m=1}^{M} 
j^{\rho(0)}_{1,\varepsilon,m}(\br_\mathrm{T})\right]^2 \ ,
\end{equation}
where $\kF$ is such that the conductance is on the $M$\textsuperscript{th} plateau. 
The second-order correction in this case is then local, and proportional to the square 
of the total current density.

For the case of an AQPC the conductance correction on the $M$\textsuperscript{th} plateau 
takes the simple form
\begin{equation}
\label{dg_vs_current2_AQPC}
g_{\rm AQPC}^{(2)}(\br_{\rm T}) = - \left[\frac{2 \Me}{\pi w \hbar^2}  
 \frac{v_{\rm T}}{\kF \rho} \sum_{m=1}^{M} 
\frac{\Theta^2_{m}(\kF,\theta)}
{{\rm Re}\lbrace{\cal K}_{m}\rbrace}\right]^2,
\end{equation}
and is plotted in Fig.~\ref{fig_dg_plateaus}. Notice that our expression derives from 
Eq.\ \eqref{eq:deltag2plateau}. Thus we are approximating the full expression \eqref{eq:deltag2} 
for $g^{(2)}$ by taking $t_m=1$, and therefore neglecting the small reflection visible in the 
eigenmode structure of Fig.\ \ref{fig_magicmode2}.
As in Eq.\ \eqref{eq:deltag1_AQPC}, the above expression is symmetric, and valid in both regions I and II. 
All open channels contribute and the overall radial decay is as $(\kF \rho)^{-2}$, consistently with 
the quantum mechanical numerical calculations.\cite{JSTW_2010} It is important to remark that the 
simple connection \eqref{dg_vs_current2_4F} between the second-order conductance correction and the 
current density is not universal, but only holds under the very restrictive conditions discussed above. 
Even for an ideal weak probe, one has to expect deviations between the induced conductance 
change and the current density in a system that does not show conductance quantization or 
if the latter arises in a non-symmetric structure. 

While we expect that realistic QPCs can be approximately symmetric, \cite{Ensslin_symm_QPC} it is 
clear that once disorder is introduced in the 2DEG neighboring the QPC, the symmetry of the 
unperturbed structure would be completely lost on scales larger than the elastic mean-free-path.

Once the 4F symmetry is broken the connection between the two scattering eigenstates 
$\chi_{1,\varepsilon,m}$ and $\chi_{2,\varepsilon,m}$ is lost, since the matrices $u_1$ and $u_2$ of 
the polar decomposition \eqref{eq:Spol} are independent, and therefore the conductance correction 
\eqref{dg_vs_current2} depends on the probability densities associated with two different states. 
Moreover, in the case where there is scattering between the QPC and the tip, the separability of the 
scattering eigenstate in the radial and angular coordinates is lost, leading to a current density which 
is no longer directed along the radial direction and proportional to the probability density [as in 
Eqs.\ \eqref{eq:jmagic} and \eqref{eq:jmagicI}].

From the above considerations we conclude that a quantitative analysis establishing the regimes for 
which the proportionality between the conductance correction, due to the presence of a weakly invasive 
tip, and the square of the current density holds, crucially depends on the symmetries of the 
unperturbed structure and on the potential landscape generated by the neighboring impurities of the 
2DEG. The relation between the measured conductance change and the current density is not universal, 
but depends on the particular SGM setup. Though a more detailed study of the scattering eigenstate 
profiles is beyond the scope of our present work, it is a relevant issue for a conclusive, clear-cut 
interpretation of SGM experiments performed in generic geometries and under different regimes. 

\section{Conclusions}
\label{sec_conclusions}

In this work we have presented the theory of scanning gate microscopy applicable when a weak probe 
is scanned above a 2DEG in the neighborhood of a nanostructure. We have discussed and extended the 
perturbative approach put forward in Ref.~[\onlinecite{JSTW_2010}] and applied it to the 
paradigmatic case in which the nanostructure is a QPC. In the case of an AQPC an approximate 
solution of the scattering eigenstates is derived and used to understand the effect of a weakly 
invasive SGM tip.

The first-order conductance correction in the tip potential involves the unperturbed scattering
amplitudes and scattering states impinging on the structure from opposite sides. 
In the case of a QPC this first order is the dominant correction only at the conductance steps
of partially open eigenmodes, but it is suppressed on conductance plateaus. This correction is 
local in space, oscillates on the scale $2k_\mathrm{F}$, and decays as the inverse of the 
distance between the QPC and the tip, but it is not directly related with the probability density 
of the scattering eigenstates, nor with their current density. When the Fermi energy is set on a 
conductance plateau, the second order correction in the tip potential becomes dominant. In the 
case of a symmetric structure it becomes a local quantity, given by the square of the local 
current density arising from all the open eigenmodes.

The dependence of the conductance correction on tip voltage has been observed to be linear in the 
case of a non-quantized conductance\cite{pala08a}, in agreement with the result \eqref{eq:deltag1} 
for $g^{(1)}$. The quadratic dependence of $g^{(2)}$ on the tip-induced potential $\VT$ is an 
important prediction that should be observable on the conductance plateaus.

Within the framework of linear response theory in the applied bias voltage $V$,
when the measured quantity is the full current across a device, the 
detailed local information encoded in the current density and the electric field is 
irrelevant.\cite{baranger91a} Only integrated quantities, the total current carried 
by the scattering states at the Fermi energy and the applied voltage, appear in the 
determination of the conductance. On the other hand, the SGM technique opens up
new possibilities, since the spatial resolution that it provides
yields conductance corrections that could be interpreted as a measure of local properties.
However, in the non-invasive case when the SGM tip acts as a weak perturbation, it is only 
under very specific conditions of symmetry and unitary propagation of the eigenmodes that 
these corrections could be shown to be local, and related to the current density.

The abrupt quantum point contact is a good model for the constrictions used in some of 
the experiments.\cite{schnez11b,Ensslin_symm_QPC} Constrictions with more gradual openings
exhibit collimation phenomena, where the momentum distribution of the electrons leaving 
the junction is weighted towards large longitudinal momentum.\cite{baranger91a}
In such cases, the current densities are more forward focused than those presented 
in Figs.\ \ref{fig_magicmode1} and \ref{fig_magicmode2}. The corresponding experimentally 
observed conductance corrections also exhibit some forward focusing.\cite{topinka00a,topinka01a} 
Even though an analytic determination of the scattering eigenstates is not possible for such 
geometries, the qualitative conclusions of our calculations remain valid, and the symmetry 
requirements necessary for establishing a connection between the conductance correction and 
the local current density also hold in this case.  

There are a number of ingredients, beyond the scope of the present paper, that might be of 
quantitative relevance in certain parameter regimes. Among them, non-perturbative effects 
in the tip potential, non-linear effects in the bias voltage, as well as interaction, 
temperature, and disorder effects.

The tip potentials currently used are strong enough to create a divot of depletion in the 
2DEG,\cite{topinka00a,topinka01a} and multiple scattering between the nanostructure and the tip.
This is not a fundamental restriction, since the matrix elements \eqref{eq:meVT} of the 
perturbation become progressively smaller as the tip moves away from the nanostructure. 
The Green function approach sketched in Appendix~\ref{app_green} is cumbersome due to the 
proliferation of energy integrals, but it is amenable to be extended to higher-order 
in the perturbation scheme.  
A complete resummation of the Born series, when possible,
should enable a connection with the non-perturbative regime of strong AFM probes. 

While most of the SGM experiments have measured linear conductances, some recent experiments 
investigated the non-linear regime.\cite{jura10} The non-linear theory of SGM is considerably more 
complicated than its linear counterpart described in this work since electron-electron interactions 
have to be incorporated at least in a self-consistent form in order to respect charge conservation. 
The spatial symmetry breaking induced by the perturbing tip might give rise, under certain conditions, 
to a measurable asymmetry of the $I$-$V$ characteristic.\cite{GWJ}

The electron-electron interactions are crucial in the non-linear regime, but they are also expected 
to be important in the linear one, in particular in the case of partially open modes.\cite{freyn08a} 
The Friedel density oscillations induced by the tip are expected to modify the electrostatic potential 
in the constriction. This effect, different from the cross talk between the tip and the gates defining 
the structure,\cite{jura09a} has been treated in one-dimensional lattice models in 
Ref.\ \onlinecite{weinmann08}.

Temperature effects are important to suppress quantum interference within the constriction and 
to improve the conductance quantization. \cite{lindelof08} When the smoothing provided by a finite 
temperature is large enough, the step and plateau behaviors get mixed, and non-monotonous 
dependencies of the fringes appearing in the conductance corrections on temperature can be 
obtained.\cite{abbout11}   

Disorder is always present in the 2DEG neighboring the nanostructure. The sample quality translates 
in the disorder strength of the 2DEG measured in terms of the mean free path, and the dominant 
scattering characteristics of short- or long-range disorder.\cite{topinka01a,jura07} 
When long-range potentials leading to small-angle scattering are dominant, the SGM sweeps reveal 
a branching structure that has been interpreted in terms of caustics related with the classically 
underlying electron motion. The presence of disorder in the 2DEG breaks any possible spatial symmetry 
of the nanostructure under study. It remains to be determined beyond which distance the symmetry 
breaking becomes relevant.

\appendix

\section{Energy integrations for the current corrections}
\label{app_integrations}

In this appendix we present the energy integrations appearing in Eqs.~\eqref{eq:deltaI} and 
\eqref{eq:deltaI2}, which are characteristic of scattering theory in 
wave-guides.\cite{FisherLee,mello04} The general form of these integrals can be expressed as
\begin{equation}
\label{integral1}
\zeta_{l,\varepsilon,a}^{\bl} = 
\int_{\tepsilon_1}^{\infty} 
\frac{\dif\bepsilon}{\varepsilon^{+}-\bepsilon} \
\sum_{\ba=1}^{\bar{N}}\, \xi^{\bl l}_{\ba a}(\bepsilon,\varepsilon) \
e^{is\bk_{\ba}^{\pm}x} \ ,
\end{equation}
where $\bk_{\ba}=
\sqrt{2\Me(\bepsilon-\tepsilon_{\ba})/\hbar^2}$. The infinitesimal positive (negative) part given to 
$\bk_{\ba}$ applies to terms arising from outgoing (incoming) modes, while $s=\pm 1$ according to 
whether the corresponding mode is a right or left-moving one. In Secs.\ \ref{sec_1st} and \ref{sec_2nd} 
we evaluate currents in the right lead, we thus are interested in the case $L/2 \ll x < \infty$, 
and the sign of $s x$ is that of $s$. In Appendix~\ref{app_green} we have the opposite case since 
$x,{\bar x} < 0$. The function $\xi^{\bl,l}_{\ba,a}(\bepsilon,\varepsilon)$ is assumed to have a 
smooth dependence on $\varepsilon$ and $\bepsilon$. 

Changing the integration variable from $\bepsilon$ to $\bk_{\ba}$ we have 
\begin{figure}
\includegraphics[width=.3\textwidth]{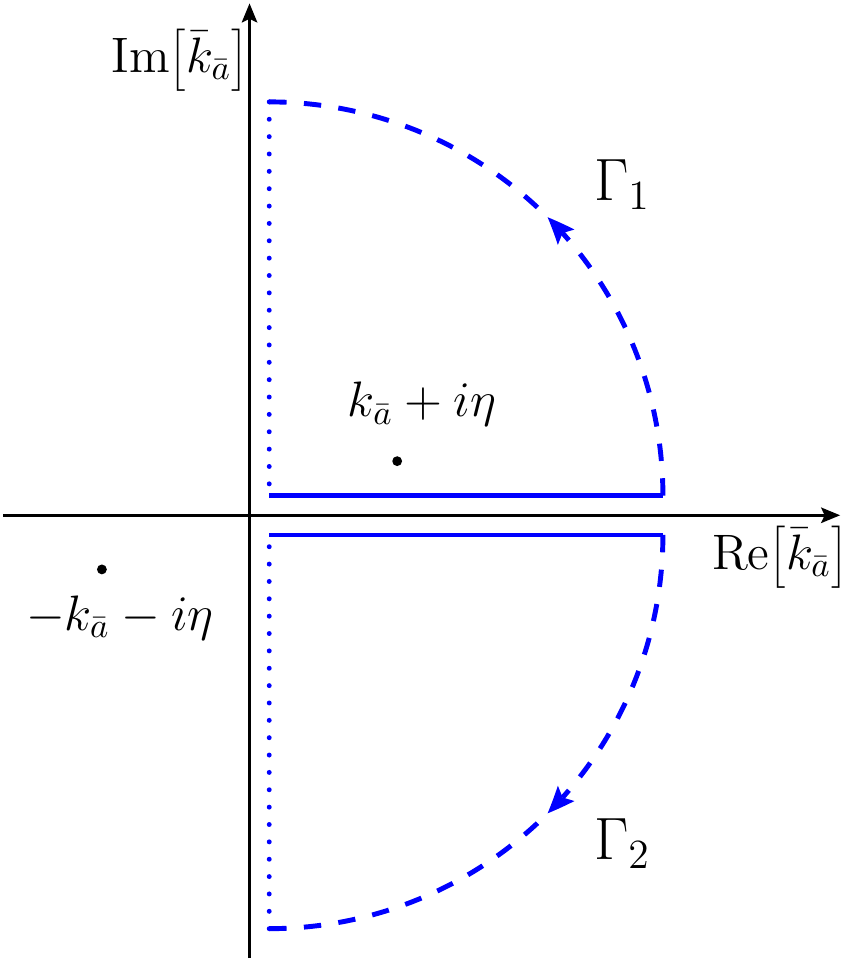}
\caption{(Color online) Contours in the complex $\bk_{\ba}$ plane for the integration of 
Eq.\ \eqref{integral2}. 
The poles shown are for an open channel $\ba$. For $sx>0\,(<0)$ the contour $\Gamma_1 (\Gamma_2)$ 
has to be chosen. This way the integral on the quarter circle (dashed lines) goes to zero for 
$\bk_{\ba}\to\infty$, whereas the one along the imaginary axis (dotted lines) becomes Laplace-like 
and thus ${\cal O}(1/|x|)$.}
\label{fig_contour}
\end{figure}
\begin{equation}
\label{integral2}
\zeta_{l,\varepsilon,a}^{\bl} = - 2 \sum_{\ba=1}^{\infty}
\int_{0}^{\infty} 
\frac{\dif\bk_{\ba}\bk_{\ba} \ \xi^{\bl l}_{\ba a}\left(\frac{\hbar^2\bk_{\ba}^2}{2\Me}+\tepsilon_{\ba},
\varepsilon\right)e^{is\bk_{\ba}^{\pm}x}}{[\bk_{\ba}-(k_{\ba}+i\eta)][\bk_{\ba}+(k_{\ba}+i\eta)]} \ , 
\end{equation}
where $k_{\ba}=\sqrt{2\Me(\varepsilon-\tepsilon_{\ba})/\hbar^2}$ is real for an open channel $\ba$ and 
imaginary for a closed one. For an open channel ($\ba \le N$) the poles in the complex $\bk_{\ba}$ plane 
are indicated in Fig.~\ref{fig_contour}. The integral along the positive real axis can be obtained from 
the contour integration along the paths $\Gamma_1$ and $\Gamma_2$. The choice of the appropriate contour 
depends on the sign of $s$ in order to ensure that (i) the contribution from the quarter circle (dashed 
lines) vanishes exponentially in the limit $|\bk_{\ba}| \to \infty$; and (ii) the contribution along 
the imaginary axis (dotted lines) is asymptotically small, ${\cal O}(1/|x|)$ in the limit 
$x \to + \infty$. For $sx>0$, $\Gamma_1$ has to be used, and the integral is given by the residue at 
the pole $\bk_{\ba}=k_{\ba}+i\eta$. For $sx<0$, $\Gamma_2$ has to be used, leading to a 
vanishing integral. For a closed channel ($\ba > N$) the poles are close to the imaginary axis and 
both contours lead to a vanishingly small contribution. We then write   
\begin{equation}
\label{integral3}
\zeta_{l,\varepsilon,a}^{\bl} = - 2 \pi i \sum_{\ba=1}^{N}
\xi^{\bl l}_{\ba a}\left(\varepsilon,\varepsilon\right)e^{ik_{\ba}^{\pm}x} \ . 
\end{equation}
The first term of Eq.\ \eqref{eq:Z_defb} has the form above described, and the corresponding 
$\bepsilon$ integration leads to a vanishing contribution. The contribution from \eqref{eq:Z_defa} 
and the second term of Eq.\ \eqref{eq:Z_defb} are slightly different, since they can be written in 
the form 
\begin{equation}
\label{integral4}
{\tilde \zeta}_{l,\varepsilon,a}^{\bl} = \int_{\tepsilon_1}^{\infty} 
\frac{\dif\bepsilon}{\varepsilon^{+}-\bepsilon} \
\sum_{\ba=1}^{\bar{N}} \sum_{b=1}^{\hat{N}} {\tilde \xi}^{\bl l}_{\ba a}(\bepsilon,\varepsilon,b) \
e^{is\bk_{b}^{\pm}x} \ ,
\end{equation}
with $\hat{N}= \mathrm{min}\lbrace N,\bar{N}\rbrace$. Now, it is convenient to pass from the variable 
$\bk_{\ba}$ to $\bk_{b}=\sqrt{\bk_{\ba}^2+2\Me(\tepsilon_{\ba}-\tepsilon_{b})/\hbar^2}$ and write
\begin{equation}
\label{integral5}
{\tilde \zeta}_{l,\varepsilon,a}^{\bl} = - 2 \sum_{\ba=1}^{\infty} \sum_{b=1}^{N}
\int_{\bk_b^{\rm min}}^{\infty} 
\frac{\dif\bk_{b}\bk_{b} \ {\tilde \xi}^{\bl l}_{\ba a}\left(\frac{\hbar^2\bk_{b}^2}{2\Me}+\tepsilon_{b},
\varepsilon,b \right) e^{is\bk_{b}^{\pm}x}}{[\bk_{b}-(k_{b}+i\eta)][\bk_{b}+(k_{b}+i\eta)]} \ , 
\end{equation}
where $k_{b}=\sqrt{2\Me(\varepsilon-\tepsilon_{b})/\hbar^2}$ is always real since $b \le N$. 
The lower integration limit $\bk_b^{\rm min}$ is equal to $0$ if $\ba \le b$, and 
$\sqrt{2\Me(\tepsilon_{\ba}-\tepsilon_{b})/\hbar^2}$ if $\ba > b$. 
In the first case (only occurring when $\ba \le N$) the pole at $\bk_{b}=k_{b}+i\eta$ is inside 
the contour $\Gamma_1$ of the complex $\bk_{b}$ plane. In the second case the contour $\Gamma_1$ 
has its vertical component with ${\rm Re}\lbrace \bk_{b} \rbrace =\bk_b^{\rm min}$ and the pole 
is within $\Gamma_1$ only if $\tepsilon_{\ba} \le \varepsilon$, that is, if $\ba \le N$. 
The resulting non-vanishing contributions to Eq.\ \eqref{integral4} are
\begin{equation}
\label{integral6}
{\tilde \zeta}_{l,\varepsilon,a}^{\bl} = - 2 \pi i \sum_{\ba=1}^{N} \sum_{b=1}^{N}
{\tilde \xi}^{\bl l}_{\ba a}\left(\varepsilon,\varepsilon,b\right)e^{isk_{b}^{\pm}x} \ . 
\end{equation}
  
The application of Eq.\ \eqref{integral6} to the $\bepsilon$ integration of Eq.~\eqref{eq:deltaI} 
readily leads to Eqs.~\eqref{eq:I1eame}-\eqref{eq:I1ea0}.

\section{First-order conductance corrections from functional derivatives}
\label{app_ccfd}

In this appendix we show that the first-order correction $g^{(1)}$ \eqref{eq:deltag1} contains 
as a particular case, the conductance sensitivity (obtained in Ref.\ \onlinecite{gasparian96}) to a 
change of the scattering potential $U(x)$ defining a one-dimensional structure. One-dimensional systems 
are particularly simple within our approach because all the scattering states are scattering eigenstates.

Writing the dimensionless conductance as $g=|t_{\rm t}|^2$, where $t_{\rm t}$ is the total transmission 
amplitude taking into account the unperturbed scatterer and the perturbation, the variation of the 
conductance when the scattering potential is changed from $U(x)$ to $U(x)+\delta U(x)$ is
\begin{equation}\label{eq:g1inteta}
\delta g=
- 4 \pi \int \mathrm{d}x\, \eta(x) \delta U(x) \ .
\end{equation}
The function $\eta$ is, up to a multiplicative constant, the functional derivative of the conductance with respect to the electrostatic potential, and can be written as
\begin{equation}
\eta(x) = -\frac{1}{4\pi}\left(t_\mathrm{t}^*\frac{\delta t_\mathrm{t}}{\delta U(x)}+
\frac{\delta t_\mathrm{t}^*}{\delta U(x)}t_\mathrm{t}\right) \ . 
\end{equation}

The case where the unperturbed scattering potential is a delta-function $U(x)=u \delta(x)$ allows for an explicit 
evaluation of the functional derivatives. The unperturbed transmission probability 
for a particle at the Fermi energy is given by $T=1/(1+{\tilde u}^2)$ with 
${\tilde u}=u/(\hbar v_\mathrm{F})$, where $v_\mathrm{F}$ is the Fermi velocity. 
The sensitivity $\eta(x)$ is given by \cite{gasparian96}
\begin{equation}\label{eq:eta1d}
\eta(x)=\frac{\tilde u}{hv_\mathrm{F}} T^2 \left[\cos(2k_\mathrm{F}x)+
{\tilde u}\sin(2k_\mathrm{F}|x|)\right] \, .
\end{equation}    

We now approach the problem from the perturbative expression \eqref{eq:deltag1} that for a 
one-dimensional system under a change $\delta U(x)$ of the scattering potential yields the 
first-order conductance correction
\begin{equation}\label{eq:1dpertg1}
g^{(1)}=-4\pi \mathrm{Im}\left\{r^*t'\int\mathrm{d}x\, 
\Psi_2^{(0)*}(x) \ \delta U(x) \Psi_1^{(0)}(x)\right\}\, ,
\end{equation}
with all quantities taken at the Fermi energy. The transmission and reflection amplitudes of the 
unperturbed delta-barrier potential are $t=t'=1/(1+i{\tilde u})$ and 
$r=r'=-i{\tilde u}/(1+i{\tilde u})$, respectively. Therefore,
\begin{equation}\label{eq:1dpertg1delta}
g^{(1)}=-4\pi {\tilde u} T\int\mathrm{d}x\, 
\mathrm{Re}\left\{\Psi_2^{(0)*}(x)\Psi_1^{(0)}(x)\right\}\delta U(x) \, .
\end{equation}
The expressions \eqref{allscats} of the lead states  yield
\begin{equation}\label{eq:rep2p1}
\mathrm{Re}\left\{\Psi_2^{(0)*}(x)\Psi_1^{(0)}(x)\right\}=\frac{T}{hv_\mathrm{F}}\left[
\cos(2k_\mathrm{F}x)+{\tilde u}\sin(2k_\mathrm{F}|x|)\right]\, .
\end{equation}
By plugging Eq.\ \eqref{eq:rep2p1} in Eq.\ \eqref{eq:1dpertg1delta}, one can see that the 
first-order conductance correction obtained from our perturbative approach can be written 
in the form Eq.\ \eqref{eq:g1inteta}, with a sensitivity $\eta(x)$ that coincides with the 
result \eqref{eq:eta1d} from the functional derivative approach of 
Ref.\ [\onlinecite{gasparian96}].

\section{Second-order correction to the current carried by a scattering state}
\label{app_secondorder}

In this appendix we provide the details of the calculation of the second-order correction, in the tip 
potential $V_T$, of the current carried by the scattering state $1,\varepsilon,a$. The two contributions
$j_{l,\varepsilon,a}^{x(2)\gamma}(\br)$, with $\gamma=\alpha,\beta$, of the second-order current 
density corrections \eqref{eq:j2}, when integrated over a cross section $\mathcal{S}_x$ at the right 
of the scatterer, lead to the corresponding changes of the current 
\begin{widetext}
\begin{subequations}
\label{eq:deltaI2}
\begin{eqnarray}
I_{1,\varepsilon,a}^{(2)\alpha}(x) &=& \frac{e\hbar}{M} \ \sum_{\bl,\bbl=1}^{2}
\mathrm{Re}\left\lbrace
\int_{\tepsilon_1}^{\infty} 
\frac{\dif\bepsilon}{\varepsilon^{+}-\bepsilon} \
\int_{\tepsilon_1}^{\infty} 
\frac{\dif\bbepsilon}{\varepsilon^{+}-\bbepsilon} \
\sum_{\ba=1}^{\bar{N}} \sum_{\bba=1}^{\bar {\bar{N}}} 
Z^{1 \bl}_{a,\ba}(\varepsilon,\bepsilon) \ 
\left[\VT\right]^{\bl \bbl}_{\ba \bba}(\bepsilon,\bbepsilon)
\left[\VT\right]^{\bbl 1}_{\bba a}(\bbepsilon,\varepsilon)
\right\rbrace  \, ,
\\
I_{1,\varepsilon,a}^{(2)\beta}(x) &=& \frac{e\hbar}{M} \ \sum_{\bl,\bbl=1}^{2}
\mathrm{Re}\left\lbrace
\int_{\tepsilon_1}^{\infty} 
\frac{\dif\bepsilon}{\varepsilon^{-}-\bepsilon} \
\int_{\tepsilon_1}^{\infty} 
\frac{\dif\bbepsilon}{\varepsilon^{+}-\bbepsilon} \
\sum_{\ba=1}^{\bar{N}} \sum_{\bba=1}^{\bar {\bar{N}}} 
\left[\VT\right]^{1 \bl}_{a \ba}(\varepsilon,\bepsilon)
X^{\bl \bbl}_{\ba \bba}(\bepsilon,\bbepsilon) 
\left[\VT\right]^{\bbl 1}_{\bba a}(\bbepsilon,\varepsilon)
\right\rbrace  \, .
\end{eqnarray}
\end{subequations}
Here, $Z^{1 \bl}_{a \ba}(\varepsilon,\bepsilon)$ is defined in
(\ref{eq:allZ_def}) and
\begin{subequations}
\begin{eqnarray}
\label{eq:X_def}
X^{11}_{a \ba}(\varepsilon,\bepsilon) &=& c^2  
\sum_{b=1}^{\hat{N}} \sqrt{\frac{\bk_b}{k_b}} \
t^{*}_{ba} {\bar t}^{\phantom{*}}_{b \ba} \
\exp{\left[i(\bk_{b}^{+}-k_{b}^{-})x\right] } \, ,  \\
X^{12}_{a \ba}(\varepsilon,\bepsilon) &=& c^2
\left\lbrace -
\sqrt{\frac{\bk_{\ba}}{k_{\ba}}} \ t^{*}_{\ba a} \
\exp{\left[-i(\bk^{-}_{\ba}+k^{-}_{\ba})x\right ] }
+ \sum_{b=1}^{\hat{N}} \sqrt{\frac{\bk_b}{k_b}} \ t^{*}_{ba}{\bar r}'_{b\ba} \
\exp{\left[i(\bk^{+}_{b}-k^-_b)x\right]}
\right\rbrace , \\
X^{21}_{a \ba}(\varepsilon,\bepsilon) &=& c^2
\left\lbrace 
\sqrt{\frac{\bk_{a}}{k_{a}}} \ {\bar t}^{*}_{a \ba} \
\exp{\left[i(\bk^{+}_{a}+k^{+}_{a})x\right ] }
+ \sum_{b=1}^{\hat{N}} \sqrt{\frac{\bk_b}{k_b}} \ r'^{*}_{ba} {\bar t}_{b \ba} \
\exp{\left[i(\bk^{+}_{b}-k^-_b)x\right]}
\right\rbrace , \\
X^{22}_{a \ba}(\varepsilon,\bepsilon) &=& c^2
\left\lbrace
-\delta_{a \ba} \ \sqrt{\frac{\bk_{a}}{k_{a}}} \
\exp{\left[-i(\bk^{-}_{\ba}-k^{+}_{a})x\right ] } +
\sqrt{\frac{\bk_{a}}{k_{a}}} \ {\bar r}'_{a \ba}
\exp{\left[i(\bk^{+}_{a}+k^{+}_{a})x\right ] } 
\right. \nonumber
\\
& & 
-\left.
\sqrt{\frac{\bk_{\ba}}{k_{a}}} \ r'^{*}_{\ba a} \
\exp{\left[-i(\bk^{-}_{\ba}+k^{-}_{\ba})x\right ] } + 
\sum_{b=1}^{\hat{N}} \sqrt{\frac{\bk_b}{k_b}} \ r'^{*}_{ba} {\bar r}'_{b \ba} \
\exp{\left[i(\bk^{+}_{b}-k^-_b)x\right]}
\right\rbrace \ ,
\end{eqnarray}
\end{subequations}
\end{widetext}
where the barred longitudinal wave-vectors and scattering amplitudes are taken at the total energy 
$\bepsilon$, and $\hat{N}= \mathrm{min}\lbrace N,\bar{N}\rbrace$. 

For the current correction 
$I_{1,\varepsilon,a}^{(2)\alpha}(x)$, only one of the
energy integrations can be done as in Appendix \ref{app_integrations}. For 
$I_{1,\varepsilon,a}^{(2)\beta}(x)$ the $\bepsilon'$ integration can be done as in Appendix 
\ref{app_integrations}, while the $\bepsilon$ integration is analogous, up to the fact that the 
poles are in the second and fourth quadrants of the complex plane for $\bk_{\ba}$ and $\bk_{b}$. 
We then have
\begin{equation}
\label{eq:I2alphaeame}
I_{1,\varepsilon,a}^{(2)\gamma}  = \l[{\cal I}_{1,\varepsilon}^{(2)\gamma}\r]_{a,a} \, , 
\end{equation}
with $\gamma = \alpha,\beta$ and the $N \times N$ matrices
\begin{widetext}
\begin{subequations}
\begin{eqnarray}
\label{eq:I2alphaea}
{\cal I}_{1,\varepsilon}^{(2)\alpha} &=& 
\frac{2e}{\hbar}\sum_{\bl=1}^{2}
\mathrm{Im}\left\lbrace
\int_{\tepsilon_1}^{\infty} 
\frac{\dif\bepsilon}{\varepsilon^+-\bepsilon} \
\left(
t^{\dagger}t \mathcal{V}^{1 \bl}(\varepsilon,\bepsilon)
\mathcal{V}^{\bl 1}(\bepsilon,\varepsilon) +
t^{\dagger}r' \mathcal{V}^{2 \bl}(\varepsilon,\bepsilon)
\mathcal{V}^{\bl 1}(\bepsilon,\varepsilon) \right)
\right\rbrace 
\\
\label{eq:I2betaea}
{\cal I}_{1,\varepsilon}^{(2)\beta} &=& \frac{e}
{2 \pi \hbar} \mathrm{Re}\left\lbrace
\mathcal{V}^{11}(\varepsilon,\varepsilon)
t^{\dagger}t \mathcal{V}^{11}(\varepsilon,\varepsilon) +
\mathcal{V}^{12}(\varepsilon,\varepsilon)
r'^{\dagger}r' \mathcal{V}^{21}(\varepsilon,\varepsilon) +
2 \mathcal{V}^{11}(\varepsilon,\varepsilon)
t^{\dagger}r' \mathcal{V}^{21}(\varepsilon,\varepsilon)
\right\rbrace \, .
\end{eqnarray} 
\end{subequations}
\end{widetext}
The corrections \eqref{eq:I2alphaeame}, summed over $a$, lead to 
the second-order correction $g^{(2)}$ given in Eq.\ \eqref{eq:deltag2alt} 
of Sec.~\ref{sec_2nd}.

\section{Green function approach to conductance  corrections}
\label{app_green}

In this appendix we provide an alternative derivation of our main results 
Eqs.\ \eqref{eq:deltag1} and \eqref{eq:deltag2alt} 
based on the perturbative expansion of the total Green function. Dyson's equation
\begin{equation}\label{eq:Dyson}
\mathcal{G}(\br,\barbr,\varepsilon)=
\mathcal{G}^{(0)}(\br,\barbr,\varepsilon)+
\int \dif \barbarbr \ 
\mathcal{G}^{(0)}(\br,\barbarbr,\varepsilon)
\VT(\barbarbr)
\mathcal{G}(\barbarbr,\barbr,\varepsilon) 
\ 
\end{equation}
yields an equivalent approach, with respect to the Lippmann-Schwinger equation 
\eqref{eq:Lippmann-Schwinger}, to the perturbative treatment of the scattering problem.

The Landauer-B\"uttiker equation \eqref{eq:g0} for the unperturbed problem and the relationship 
\eqref{allTRAMs} between the transmission amplitudes and the Green function, can also be used 
to relate the total conductance $g$ with the total transmission amplitude $t_\mathrm{t}$, and 
the total scattering amplitudes in terms of the total Green function $\mathcal{G}$. Going to 
second order in $\VT$ in the expansion of the total reflection amplitude 
$r_\mathrm{t} \simeq r + \delta r^{(1)} + \delta r^{(2)}$ we have
\begin{subequations}
\label{eq:deltag1a2}
\begin{eqnarray}
\label{eq:deltag1a2seq1}
\delta g^{(1)} &=& -2 \mathrm{Re}\left\lbrace \mathrm{Tr}
\left[r^{\dagger}\delta r^{(1)}
\right]\right\rbrace
\, ,
\\
\delta g^{(2)} &=& \delta g^{(2)\alpha} + 
\delta g^{(2)\beta}  \, ,
\end{eqnarray}
\end{subequations}
with
\begin{subequations}
\label{eq:deltag2alphabeta}
\begin{eqnarray}
\delta g^{(2)\alpha} &=& -2 \mathrm{Re}\left\lbrace \mathrm{Tr}
\left[r^{\dagger}\delta r^{(2)}
\right]\right\rbrace
\, ,
\\
\delta g^{(2)\beta} &=& - \mathrm{Re}\left\lbrace \mathrm{Tr} \left[\delta r^{(1)\dagger}\delta r^{(1)}
\right]\right\rbrace  \, ,
\end{eqnarray}
\end{subequations}
where the energy arguments are taken at $\EF$.
Using the first-order approximation of Eq.\ \eqref{eq:Dyson} one gets
\begin{widetext}
\begin{equation}
\label{eq:deltar1}
\delta r_{ba}^{(1)} = i \hbar(v_{a}v_{b})^{1/2}
\ \exp{\left[-i(k_b^{+} x + k_a^{+} {\bar x})\right]} 
\int_{\SC_{x}} \dif y 
\int_{\SC_{\bar x}} \dif {\bar y} 
\int \dif \barbarbr
\ \phi_{b}^{*}(y) \ 
\mathcal{G}^{(0)}(\br,\barbarbr,\varepsilon) 
\ \VT(\barbarbr) \
\mathcal{G}^{(0)}(\barbarbr,\barbr,\varepsilon)
\ \phi_{a}({\bar y}) \ .
\end{equation}

From the spectral decomposition \eqref{eq:G0} of $\mathcal{G}^{(0)}$ in the basis of 
scattering states $\Psi_{l,\varepsilon,a}^{(0)}$, the definition \eqref{eq:meVT} of the matrix 
elements of the perturbation, and integrating over the transverse coordinates $y$ and ${\bar y}$ we have
\begin{eqnarray}
\label{eq:delta_r_1}
\delta r_{ba}^{(1)} &=& 
\frac{i (v_{a}v_{b})^{1/2}}{2 \pi} \ 
\ \exp{\left[-i(k_b^{+} x + k_a^{+} {\bar x})\right]}
\int_{\tepsilon_1}^{\infty} 
\frac{\dif\bbepsilon}{\varepsilon^{+}-\bbepsilon} \ 
\int_{\tepsilon_1}^{\infty} 
\frac{\dif\bepsilon}{\varepsilon^{+}-\bepsilon} \ 
\frac{1}{({\bar {\bar v}}_{a}{\bar v}_{b})^{1/2}}
\nonumber \\ & &
\sum_{\ba,\bba=1}^{N}
\left\lbrace
\left(\delta_{a \bba}
\exp{\left[-i\bbk^{+}_{a}{\bar x}\right ]} +
{\bar {\bar r}}^{*}_{a \bba}
\exp{\left[i\bbk^{-}_{a}{\bar x}\right ]}\right)
\left( \delta_{b \ba}
\exp{\left[i\bk^{-}_{b}x\right]} +
{\bar r}_{b \ba}
\exp{\left[-i\bk^{+}_{b} x\right ]} \right)
\left[\VT\right]^{11}_{\ba\bba}(\bepsilon,\bbepsilon) \ +
\right.
\nonumber \\ & &
\left(\delta_{a \bba}
\exp{\left[-i\bbk^{+}_{a}{\bar x}\right ]} +
{\bar {\bar r}}^{*}_{a \bba}
\exp{\left[i\bbk^{-}_{a}{\bar x}\right ]}\right)
{\bar t}_{b \ba}^{\prime}
\exp{\left[-i\bk^{+}_{b} x\right ]}
\left[\VT\right]^{21}_{\ba\bba}(\bepsilon,\bbepsilon) \ +
\nonumber \\ & &
{\bar t}^{\prime *}_{a \bba}
\exp{\left[i\bbk^{-}_{a}{\bar x}\right ]}
\left( \delta_{b \ba}
\exp{\left[i\bk^{-}_{b}x\right]} +
{\bar r}_{b \ba}
\exp{\left[-i\bk^{+}_{b} x\right ]} \right)
\left[\VT\right]^{12}_{\ba\bba}(\bepsilon,\bbepsilon) \ +
\nonumber \\ & &
\left.
{\bar t}^{\prime *}_{a \bba}
\exp{\left[i\bbk^{-}_{a}{\bar x}\right ]}
{\bar t}_{b \ba}^{\prime}
\exp{\left[-i\bk^{+}_{b} x\right ]}
\left[\VT\right]^{22}_{\ba\bba}(\bepsilon,\bbepsilon)
\right\rbrace \ .
\end{eqnarray}
\end{widetext}

Performing the $\bbepsilon$ and $\bepsilon$ integrations according to the prescription of 
Appendix~\ref{app_integrations} for the case $x,{\bar x}<0$, the previous expression reduces to 
\begin{equation}
\label{eq:delta_r_1p}
\delta r_{ba}^{(1)} = - 2\pi i 
\sum_{\ba=1}^{N}
\left\lbrace r_{b \ba}
\left[\VT\right]^{11}_{\ba a}
(\varepsilon,\varepsilon) \ +
t_{b \ba}^{\prime}
\left[\VT\right]^{21}_{\ba a}
(\varepsilon,\varepsilon) 
\right\rbrace  \, .
\end{equation}

According to Eq.\ \eqref{eq:deltag1a2seq1} 
\begin{equation}
\label{eq:deltag1GF}
g^{(1)}  = - 4 \pi\ \mathrm{Im}\left\lbrace \mathrm{Tr} \left[
r^{\dagger}r\ \mathcal{V}^{11} +
r^{\dagger}t'\ \mathcal{V}^{21}\right]\right\rbrace \, ,
\end{equation}
with the matrices $\mathcal{V}^{\bl l}$ defined as in Sec.~\ref{sec_1st} from Eq.~\eqref{eq:meVT}. As noticed in Sec.~\ref{sec_1st}, the first term gives a vanishing contribution, and therefore we recover 
Eq.~\eqref{eq:deltag1}. 

The second-order variation of the reflection amplitude is
\begin{widetext}
\begin{eqnarray}
\label{eq:deltar2}
\delta r_{ba}^{(2)} &=& i \hbar(v_{a}v_{b})^{1/2}
\ \exp{\left[-i(k_b^{+} x + k_a^{+} {\bar x})\right]}
\int_{\SC_{x}} \dif y 
\int_{\SC_{\bar x}} \dif {\bar y} 
\int \dif \barbarbr^{\prime} \int \dif \barbarbr
\nonumber \\
& & 
\ \phi_{b}^{*}(y) \ 
\mathcal{G}^{(0)}(\br,\barbarbr^{\prime},\varepsilon) 
\ \VT(\barbarbr^{\prime}) \
\mathcal{G}^{(0)}(\barbarbr^{\prime},\barbarbr,\varepsilon) 
\ \VT(\barbarbr) \
\mathcal{G}^{(0)}(\barbarbr,\barbr,\varepsilon)
\ \phi_{a}({\bar y}) \ .
\end{eqnarray}

The structure of \eqref{eq:deltar2} is rather similar to that of \eqref{eq:deltar1} for the correction
$\delta r_{ba}^{(1)}$. The main difference is that the spectral decomposition of the unperturbed Green function now leads to three energy integrations, where two of them can be done as before, yielding
\begin{equation}
\label{eq:delta_r_2p}
\delta r_{ba}^{(2)} = - 2\pi i \sum_{\bl=1}^{2}
\int_{\tepsilon_1}^{\infty} 
\frac{\dif\bepsilon}{\varepsilon^{+}-\bepsilon} \ 
\sum_{\ba,\bba=1}^{N}
\left\lbrace
r_{b \bba}
\left[\VT\right]^{1\bl}_{\bba\ba}(\varepsilon,\bepsilon) 
\left[\VT\right]^{\bl 1}_{\ba a}(\bepsilon,\varepsilon) 
\ +
t_{b \bba}^{\prime}
\left[\VT\right]^{2 \bl}_{\bba\ba}(\varepsilon,\bepsilon) 
\left[\VT\right]^{\bl 1}_{\ba a}(\bepsilon,\varepsilon) 
\right\rbrace
\, .
\end{equation}
Then,
\begin{equation}
\label{eq:deltag2alphaGF}
g^{(2)\alpha}  = - 4 \pi\ \sum_{\bl=1}^{2}
\left(\pi \mathrm{Tr} \left[
r^{\dagger}r\ \mathcal{V}^{1 \bl} \mathcal{V}^{\bl 1} \right] +
\mathrm{Im}\left\lbrace 
\int_{\tepsilon_1}^{\infty} 
\frac{\dif\bepsilon}{\bepsilon-\EF^{+}} \ 
\mathrm{Tr} \left[r^{\dagger}t'\ 
\mathcal{V}^{2 \bl}(\EF,\bepsilon) \mathcal{V}^{\bl 1}(\bepsilon,\EF) \right]\right\rbrace
\right) \, ,
\end{equation}
where the non-specified energy arguments are understood to be taken at $\EF$. Since
\begin{equation}
\label{eq:deltag2betaGF}
g^{(2)\beta}  = - 4 \pi^2 \mathrm{Tr} 
\left[r^{\dagger}r \ \mathcal{V}^{11} \mathcal{V}^{11} +
t^{\prime \dagger}t^{\prime} \ \mathcal{V}^{21} \mathcal{V}^{12} +
2 \mathrm{Re}\left\lbrace 
r^{\dagger}t' \ \mathcal{V}^{21} \mathcal{V}^{11}
\right\rbrace\right] 
\, ,
\end{equation}
the second-order conductance correction can be written as
\begin{eqnarray}
g^{(2)}  & = & - 4 \pi^2\ \left(  
\mathrm{Tr}\left[t^{\prime \dagger}t^{\prime} \ \mathcal{V}^{21}\mathcal{V}^{12}
- r^{\dagger}r\ \mathcal{V}^{12}\mathcal{V}^{21}\right] -
2 \mathrm{Re}\left\lbrace \mathrm{Tr}\left[r^{\dagger}t' \mathcal{V}^{21}\mathcal{V}^{11}
\right] \right\rbrace \right)
\nonumber \\
&-& 4 \pi \sum_{\bl=1}^{2} \mathrm{Im}\left\lbrace
\int_{\tepsilon_1}^{\infty} 
\frac{\dif\bepsilon}{\bepsilon-\EF^{+}} \
\mathrm{Tr}[r^{\dagger}t'
\mathcal{V}^{2 \bl}(\EF,\bepsilon)
\mathcal{V}^{\bl 1}(\bepsilon,\EF)]
\right\rbrace
\, . 
\label{eq:deltag2GF}
\end{eqnarray}
\end{widetext}

Using the unitarity condition for $S$ we see that Eq.\ \eqref{eq:deltag2GF} is equivalent to 
Eq.\ \eqref{eq:deltag2alt}. The symmetry between the two ways of writing $g^{(2)}$ arises from
the different choice of the cross sections ($x,{\bar x}<0$ in this appendix, and $x>0$ in 
Sec.~\ref{sec_2nd}).

\section{Exact results in one dimensional models}
\label{app_verification}

In this appendix we treat the exactly solvable case of a one-dimensional scatterer perturbed by a 
local tip, and we compare the resulting conductance corrections with those predicted by 
Eqs.~(\ref{eq:deltag1}) and (\ref{eq:deltag2}). We consider a scatterer described by a 
$2 \times 2$ $S$ matrix, with the form (\ref{eq:scatt_mat}), under a perturbation
\begin{equation}
\label{eq:dfp}
V_{\rm T}(x)=v_{\rm T} \delta(x-x_{\rm T}) \ ,
\end{equation} 
where $x_{\rm T}$ is the distance between the scatterer and the tip. The dimensionless conductance 
of the combined one-dimensional system is
\begin{equation}
\label{eq:deltagtodm}
g_{1 \rm d} = \frac{1}{1- 2\tilde{v} \ \mathrm{Im}\left\lbrace r' e^{i \alpha} \right\rbrace +
\tilde{v}^2\left(1+|r'|^2+2\mathrm{Re}\left\lbrace r' e^{i \alpha} \right\rbrace \right) }\, .  
\end{equation}
We note $\alpha=2k x_{\rm T}$ and $\tilde{v}=v_{\rm T}/\hbar v$. The first and second order 
conductance corrections should then be
\begin{subequations}
\label{allcorrect}
\begin{equation}
\label{correct1}
g_{1 \rm d}^{(1)} = 2\tilde{v} \ |t|^2 \ \mathrm{Im}\left\lbrace r' e^{i \alpha} \right\rbrace \, .  
\end{equation}
\begin{equation}
\label{correct2}
g_{1 \rm d}^{(2)} = - \tilde{v}^2 \ |t|^2 \left(1+|r'|^2+2\mathrm{Re}\left\lbrace r' 
e^{i \alpha} \right\rbrace ) 
- 4 \left[\mathrm{Im}\left\lbrace r' e^{i \alpha} \right\rbrace \right]^2 \right) \, .  
\end{equation}
\end{subequations}

In order to make the connection with Eqs.~(\ref{eq:deltag1}) and (\ref{eq:deltag2}) the
matrix elements (\ref{eq:meVT}) involving the unperturbed scattering states and the tip 
potential (\ref{eq:dfp}) need to be evaluated. This is easily done in the one-dimensional 
case that we are treating
\begin{widetext} 
\begin{subequations}
\label{eq:matelodm}
\begin{eqnarray}
\label{eq:matelodma}
\mathcal{V}^{11}(\bepsilon,\varepsilon) &=& 
\frac{\tilde{v}}{2\pi} \sqrt{\frac{k}{\bk}} \
{\bar t}^{*} t \ \exp{\left[i(k-\bk)x_{\rm T}\right] } \, ,  \\
\mathcal{V}^{21}(\bepsilon,\varepsilon) &=& 
\frac{\tilde{v}}{2\pi} \sqrt{\frac{k}{\bk}} \ 
t \ \exp{\left[i k x_{\rm T}\right] }  
\left\lbrace \exp{\left[i \bk x_{\rm T} \right]} + {\bar r}'^{*} \ \exp{\left[-i \bk x_{\rm T}\right]} 
\right\rbrace \, , \\
\mathcal{V}^{22}(\bepsilon,\varepsilon) &=& 
\frac{\tilde{v}}{2\pi} \sqrt{\frac{k}{\bk}} \
\left\lbrace \exp{\left[i \bk x_{\rm T} \right]} + {\bar r}'^{*} \ \exp{\left[-i \bk x_{\rm T}\right]} 
\right\rbrace
\left\lbrace \exp{\left[- i k x_{\rm T} \right]} + r' \ \exp{\left[i k x_{\rm T}\right]} 
\right\rbrace
\ .
\end{eqnarray}
\end{subequations}
The first order correction (\ref{eq:deltag1}) only involves matrix elements evaluated at the
Fermi energy $\EF$, and can be written as
\begin{equation}
\label{eq:deltag1odm}
g^{(1)}  = - 4 \pi\ \mathrm{Im}\left\lbrace r^{*}t'\ \mathcal{V}^{21} \right\rbrace 
= 2\tilde{v} \ |t|^2 \ \mathrm{Im}\left\lbrace r' e^{i k x_{\rm T}} \left( e^{i k x_{\rm T}} 
+ r^{\prime *} e^{-i k x_{\rm T}} \right)  \right\rbrace
\, ,  
\end{equation}
which is equivalent to Eq.~(\ref{correct1}). 
Using the form (\ref{eq:matelodm}) of the matrix elements, the last contribution of $g^{(2)}$ in 
Eq.~(\ref{eq:deltag2alt})
can be written as
\begin{equation}
\begin{aligned}
\label{correct2ei}
- \frac{2\tilde{v}^2}{\pi} \ |t|^2 \ \mathrm{Im}\left\lbrace \left(|r'|^2 + 
r' e^{2 i k x_{\rm T}} \right)
\int_{0}^{\infty} 
\frac{\dif\bepsilon}{\bepsilon-\varepsilon^-} \
\left(1 + \mathrm{Re} \left\lbrace {\bar r}' e^{2 i \bk x_{\rm T}} \right\rbrace \right) 
\right\rbrace \nonumber \\
= - 2 \tilde{v}^2 \ |t|^2 \left(|r'|^2 (1 + \mathrm{Re}\left\lbrace r' e^{i \alpha} \right\rbrace ) 
+ \mathrm{Re}\left\lbrace r' e^{i \alpha} \right\rbrace +
\mathrm{Re}\left\lbrace (r' e^{i \alpha})^2 \right\rbrace
\right)
\, ,  
\end{aligned}
\end{equation}
where all terms are evaluated at $\varepsilon$ in the final expression. When combined with the 
on-shell terms of Eq.\ \eqref{eq:deltag2alt} the second-order correction of the one-dimensional model 
becomes
\begin{equation}
g^{(2)} = - u^2 \ |t|^2 \left(1 - |r'|^2 + 
2\mathrm{Re}\left\lbrace r' e^{i \alpha} \right\rbrace  
+ 2 (\mathrm{Re}\left\lbrace r' e^{i \alpha} \right\rbrace)^2 
\right) \ ,
\end{equation}
which is easy to verify that coincides with Eq.\ \eqref{correct2}.
\end{widetext} 

\acknowledgments

We thank M.\ B\"uttiker and J.-L.\ Pichard for useful discussions.
Financial support from the French National Research Agency 
ANR (Project No.\ ANR-08-BLAN-0030-02), from the German Research Foundation DFG (TRR80), from the European Union within the Initial Training Network NanoCTM, and from the U.S. National Science Foundation (Grant No.~PHY-0855337) is gratefully acknowledged.


\begin{thebibliography}{99}

\bibitem{eriksson96}
%M.\ A.\ Eriksson \etal,
M.\ A.\ Eriksson, R.\ G.\ Beck, M.\ Topinka, J.\ A.\ Katine, R.\ M.\ Westervelt, 
K.\ L.\ Campman, and A.\ C.\ Gossard,
Appl.\ Phys.\ Lett.\ \textbf{69}, 671 (1996).

\bibitem{sellier11}
%H.\ Sellier \etal,
H.\ Sellier, B.\ Hackens, M.\ G.\ Pala, F.\ Martins, S.\ Baltazar, X.\ Wallart, 
L.\ Desplanque, V.\ Bayot, and S.\ Huant,
Semicond.\ Sci.\ Technol.\ \textbf{26}, 064008 (2011).

\bibitem{topinka00a}
%M.\ A.\ Topinka \etal,
M.\ A.\ Topinka, B.\ J.\ LeRoy, S.\ E.\ J.\ Shaw, E.\ J.\ Heller, R.\ M.\ Westervelt, 
K.\ D.\ Maranowski, and A.\ C.\ Gossard,
Science \textbf{289}, 2323 (2000).

\bibitem{topinka01a}
%M.\ A.\ Topinka \etal, 
M.\ A.\ Topinka, B.\ J.\ LeRoy, R.\ M.\ Westervelt, S.\ E.\ J.\ Shaw, R.\ Fleischmann, 
E.\ J.\ Heller, K.\ D.\ Maranowski, and A.\ C.\ Gossard,
Nature \textbf{410}, 183 (2001).

\bibitem{aidala07}
K.\ E.\ Aidala, R.\ E.\ Parrott, T.\ Kramer, E.\ J.\ Heller, R.\ M.\ Westervelt, 
M.\ P.\ Hanson, and A.\ C.\ Gossard,
Nature Physics \textbf{3}, 464 (2007).

\bibitem{jura07}
M.\ P.\ Jura, M.\ A.\ Topinka, L.\ Urban, A.\ Yazdani, H.\ Shtrickman, L.\ N.\ Pfeiffer,
K.\ W.\ West, and D.\ Goldhaber-Gordon,
Nature Physics \textbf{3}, 841 (2007).

\bibitem{jura09a}
M.\ P.\ Jura, 
M.\ A.\ Topinka, M.\ Grobis, L.\ N.\ Pfeiffer, K.\ W.\ West, and D.\ Goldhaber-Gordon, 
Phys.\ Rev.\ B \textbf{80}, 041303(R) (2009).

\bibitem{jura10}
M.\ P.\ Jura, 
M.\ Grobis, M.\ A.\ Topinka, L.\ N.\ Pfeiffer, K.\ W.\ West, and D.\ Goldhaber-Gordon,
Phys.\ Rev.\ B \textbf{82}, 155328 (2010).

\bibitem{schnez11b}
%S.\ Schnez \etal,
S.\ Schnez, C.\ R\"ossler, T.\ Ihn, K.\ Ensslin, C.\ Reichl, and W.\ Wegscheider,
Phys.\ Rev.\ B \textbf{84}, 195322 (2011).

\bibitem{aoki12}
N.\ Aoki, R.\ Brunner, A.\ M.\ Burke, R.\ Akis, R.\ Meisels, D.\ K.\ Ferry, and Y.\ Ochiai,
Phys.\ Rev.\ Lett.\ \textbf{108}, 136804 (2012).

\bibitem{kozikov12}
A.\ A.\ Kozikov, C.\ R\"ossler, T.\ Ihn, K.\ Ensslin, C.\ Reichl, and W.\ Wegscheider,
New J.\ Phys.\ \textbf{15}, 013056 (2013).

\bibitem{bachtold00}
%A.\ Bachtold \etal,
A.\ Bachtold, M.\ S.\ Fuhrer, S.\ Plyasunov, M.\ Forero, E.\ H.\ Anderson, 
A.\ Zettl, and P.\ L.\ McEuen,
Phys.\ Rev.\ Lett.\ \textbf{84}, 6082 (2000).

\bibitem{pioda04}
A.\ Pioda, S.\ Ki\v{c}in, T.\ Ihn, M.\ Sigrist, A.\ Fuhrer, K.\ Ensslin, A.\ Weichselbaum, 
S.\ E.\ Ulloa, M.\ Reinwald, and W.\ Wegscheider,
Phys.\ Rev.\ Lett.\ \textbf{93}, 216801 (2004).

\bibitem{fallahi05}
%P.\ Fallahi \etal,
P.\ Fallahi, A.\ C.\ Bleszynski, R.\ M.\ Westervelt, J.\ Huang, J.\ D.\ Walls, E.\ J.\ Heller, 
M.\ Hanson, and A.\ C.\ Gossard,
Nano Lett.\ \textbf{5}, 223 (2005).

\bibitem{woodside02}
M.\ T.\ Woodside and P.\ L.\ McEuen,
Science \textbf{296}, 1098 (2002).

\bibitem{bleszynski07}
%A.\ C.\ Bleszynski \etal,
A.\ C.\ Bleszynski, F.\ A.\ Zwanenburg, R.\ M.\ Westervelt, A.\ L.\ Roest, 
E.\ P.\ A.\ M.\ Bakkers, and L.\ P.\ Kouwenhoven,
Nano Lett.\ \textbf{7}, 2559 (2007).

\bibitem{schnez11a}
%S.\ Schnez \etal,
S.\ Schnez, J.\ G\"uttinger, C.\ Stampfer, K.\ Ensslin, and T.\ Ihn,
New J.\ Phys.\ \textbf{13}, 053013 (2011).

\bibitem{hackens06}
B.\ Hackens, F. Martins, T. Ouisse, H. Sellier, S. Bollaert, X. Wallart, 
A. Cappy, J. Chevrier, V. Bayot, and S. Huant,
Nature Phys.\ \textbf{2}, 826 (2006).

\bibitem{martins07a}
F.\ Martins, B.\ Hackens, M.\ G.\ Pala, T.\ Ouisse, H.\ Sellier, X.\ Wallart, S.\ Bollaert,
A.\ Cappy, J.\ Chevrier, V.\ Bayot, and S.\ Huant, 
Phys.\ Rev.\ Lett.\ \textbf{99}, 136807 (2007).

\bibitem{pala08a}
M.\ G.\ Pala, B.\ Hackens, F.\ Martins, H.\ Sellier, V.\ Bayot, S.\ Huant, and T.\ Ouisse,
Phys.\ Rev.\ B \textbf{77}, 125310 (2008).

\bibitem{baumgartner07}
%A.\ Baumgartner \etal,
A. Baumgartner, T. Ihn, K.\ Ensslin, K.\ Maranowski, and A.\ C.\ Gossard,
Phys.\ Rev.\ B \textbf{76}, 085316 (2007).

\bibitem{paradiso10}
N.\ Paradiso, S.\ Heun, S.\ Roddaro, L.\ N.\ Pfeiffer, K.\ W.\ West, L.\ Sorba, G.\ Biasiol, 
and F.\ Beltram,
Physica E \textbf{42}, 1038 (2010).

\bibitem{paradiso11}
N. Paradiso, S.\ Heun, S.\ Roddaro, D.\ Venturelli, F.\ Taddei,
V.\ Giovannetti, R.\ Fazio, G.\ Biasiol, L.\ Sorba, and F.\ Beltram,
Phys. Rev. B \textbf{83}, 155305 (2011).

\bibitem{paradiso12}
N.\ Paradiso, S.\ Heun, S.\ Roddaro, L.\ Sorba, F.\ Beltram,
G.\ Biasiol, L.\ N.\ Pfeiffer, and K.\ W.\ West,
Phys. Rev. Lett. \textbf{108}, 246801 (2012).

\bibitem{crook03}
R.\ Crook, C.\ G.\ Smith, A.\ C.\ Graham, I.\ Farrer, H.\ E.\ Beere, and D.\ A.\ Ritchie,
Phys. Rev. Lett. \textbf{91}, 246803 (2003).

\bibitem{connolly12}
M.\ R.\ Connolly, R.\ K.\ Puddy, D.\ Logoteta, P.\ Marconcini, M.\ Roy, J.\ P.\ Griffiths,
G.\ A.\ C.\ Jones, P.\ A.\ Maksym, M.\ Macucci, and C.\ G.\ Smith,
Nano Lett. \textbf{12}, 5448 (2012).

\bibitem{Boyd_West}
E.E. Boyd and R.M. Westervelt, Phys.\ Rev.\ B \textbf{84}, 205308 (2011).

\bibitem{leroy05}
B.\ J.\ LeRoy, A.\ C.\ Bleszynski, K.\ E.\ Aidala, R.\ M.\ Westervelt, A.\ Kalben, 
E.\ J.\ Heller, S.\ E.\ J.\ Shaw, K.\ D.\ Maranowski, and A.\ C.\ Gossard,
Phys.\ Rev.\ Lett.\ \textbf{94}, 126801 (2005).

\bibitem{heller05}
E.\ J.\ Heller, K.\ E.\ Aidala, B.\ J.\ LeRoy, A.\ C.\ Bleszynski, A.\ Kalben, 
R.\ M.\ Westervelt, K.\ D.\ Maranowski, and A.\ C.\ Gossard, 
Nano Lett.\ \textbf{5}, 1285 (2005).

\bibitem{metalidis05} 
G.\ Metalidis and P.\ Bruno,
Phys.\ Rev.\ B \textbf{72}, 235304 (2005).

\bibitem{cresti06}
A.\ Cresti,
J.\ Appl.\ Phys.\ \textbf{100}, 053711 (2006).

\bibitem{kramer08}
T.\ Kramer, E.\ J.\ Heller, and R.\ E.\ Parrott,
J.\ Phys.:\ Conf.\ Ser.\ \textbf{99}, 012010 (2008). 

\bibitem{freyn08a}
A.\ Freyn, I.\ Kleftogiannis, and J.-L.\ Pichard,
Phys.\ Rev.\ Lett.\ \textbf{100}, 226802 (2008).

\bibitem{JSTW_2010}
R.A.\ Jalabert, W.\ Szewc, S.\ Tomsovic, D.\ Weinmann,
Phys.\ Rev.\ Lett.\ \textbf{105}, 166802 (2010).

\bibitem{abbout11}
A.\ Abbout, G.\ Lemari\'e, and J.-L. Pichard,
Phys.\ Rev.\ Lett.\ \textbf{106}, 156810 (2011).

\bibitem{gasparian96} V. Gasparian, T. Christen, and M.\ B\"uttiker,
Phys.\ Rev.\ A \textbf{54}, 4022 (1996).

\bibitem{FisherLee}
D.\ S.\ Fisher and P.\ A.\ Lee,
Phys.\ Rev.\ B \textbf{23}, 6851 (1981).

\bibitem{buttiker88}
M.\ B\"uttiker, IBM J.\ Res.\ Dev.\ \textbf{32}, 317 (1988).

\bibitem{jalabert00} 
R.\ A.\ Jalabert, 
in \textit{New Directions in Quantum Chaos}, ed.\ by G.\ Casati, I.\ Guarneri,
and U.\ Smilansky, IOS Press Amsterdam (2000).

\bibitem{mello04} 
P.\ A.\ Mello and N.\ Kumar, \textit{Quantum Transport in Mesoscopic Systems},
Oxford University Press, 2004.

\bibitem{LB} Y.\ Imry, {\em Introduction to Mesoscopic Systems} (Oxford
University Press, Oxford, 2002), 2nd ed.

\bibitem{corr1} The difference with Eq.~(10) in Ref.\ \onlinecite{JSTW_2010} arises from an 
erroneous simplification of the off-shell term.

\bibitem{vanwees88}
B.\ J.\ van Wees, H.\ van Houten, C.\ W.\ J.\ Beenakker, J.\ G.\ Williamson, 
L.\ P.\ Kouwenhoven, D.\ van der Marel, and C.\ T.\ Foxon,
Phys.\ Rev.\ Lett.\ \textbf{60}, 848 (1988).

\bibitem{wharam88}
D.\ A.\ Wharam, T.\ J.\ Thornton, R.\ Newbury, M.\ Pepper, H.\ Ahmed, J.\ E.\ F.\ Frost, 
D.\ G.\ Hasko, D.\ C.\ Peacock, D.\ A.\ Ritchie, and G.\ A.\ C.\ Jones,
J.\ Phys.\ C: Solid State Phys.\ \textbf{21}, L209 (1988).

\bibitem{lindelof08}
P.\ E.\ Lindelof and M.\ Aagesen,
J.\ Phys.: Condens.\ Matter \textbf{20}, 164207 (2008).

\bibitem{glazman}
L.\ I.\ Glazman, G.\ B.\ Lesovik, D.\ E.\ Khmel'nitskii, and R.\ I.\ Shekter,
Pis'ma Zh.\ Eksp.\ Teor.\ Fiz.\ \textbf{48}, 218 (1988) [JETP Lett.\
\textbf{48}, 238 (1988)].

\bibitem{szafer89}
A.\ Szafer and A.\ D.\ Stone,
Phys.\ Rev.\ Lett.\ \textbf{62}, 300 (1989).

\bibitem{kirczenow89}
G.\ Kirczenow, Phys.\ Rev.\ B \textbf{39}, 10452(R) (1989).

\bibitem{buttiker90} M.\ B\"uttiker,
Phys.\ Rev.\ B \textbf{41}, 7906(R) (1990).

\bibitem{baranger96}
H.\ U.\ Baranger and P.\ A.\ Mello,
Phys.\ Rev.\ B \textbf{54}, 14297 (1996).

\bibitem{Ensslin_symm_QPC}
C.\ R\"ossler, S. \ Baer, E. de Wiljes, P.-L. \ Ardelt, T.\ Ihn, K.\ Ensslin, 
C.\ Reichl, and W.\ Wegscheider,
New J.\ Phys.\ \textbf{13}, 113006 (2011).

\bibitem{baranger91a}
H.\ U.\ Baranger, D.\ P.\ DiVincenzo, R.\ A.\ Jalabert, and A.\ D.\ Stone,
Phys.\ Rev.\ B \textbf{44}, 10637 (1991).

\bibitem{footnote2}
In order for our perturbative approach to be sensible,
the more extended the potential, the smaller its absolute value should be.

\bibitem{GWJ}
C.\ Gorini, D.\ Weinmann, and R.\ A.\ Jalabert,
to be published.

\bibitem{weinmann08}
D.\ Weinmann, R.\ A.\ Jalabert, A.\ Freyn, G.-L.\ Ingold, and J.-L.\ Pichard,
Eur.\ Phys.\ J.\ B \textbf{66}, 239 (2008).
\end{thebibliography}
\end{document}